\documentclass[showpacs,amssymb,
eqsecnum,
twocolumn,
tightenlines,
]{revtex4}

\usepackage{bm} 

\usepackage{amsmath} 

\usepackage{dcolumn}

\begin{document}

    \bibliographystyle{apsrev}
    
    \title {Radiative corrections to parity-non-conservation in
       atoms}
    
    \author{M.Yu.Kuchiev} 
    \email[Email:]{kuchiev@newt.phys.unsw.edu.au} 
    \author{V.V.Flambaum}
    \email[Email:]{flambaum@newt.phys.unsw.edu.au} 
\affiliation{ School
      of Physics, University of New South Wales, Sydney 2052,
      Australia} 
    


    \begin{abstract} Recent progress in calculations of QED radiative
      corrections to parity nonconservation in atoms is reviewed. The
      QED vacuum polarization, the self-energy corrections and the
      vertex corrections are shown to be described very reliably by
      different methods used by different groups. All new calculations
      have recently converged to very close final values.  Each
      separate radiative correction is very large, above 1 \% for
      heavy atoms, but having different signs they partly compensate
      each other. Our results for the radiative corrections for all
      atoms are presented. The corrections are $-0.54 \% $ for
      $^{133}$Cs, and $-0.70 \% $ for $^{205}$Tl, $^{208}$Pb, and
      $^{209}$Bi.  The result for $^{133}$Cs reconciles the most
      accurate atomic experimental data for the $6s-7s$ PNC amplitude
      in $^{133}$Cs of Wood {\it et al} \cite{wood_97} with the
      standard model.

 \end{abstract}

    \pacs{32.80.Ys, 11.30.Er, 31.30.Jv}

    \maketitle

\section{Introduction}    
   \label{intro}
   
   It has become clear quite recently that the QED corrections give
   a large contribution, of the order of $\sim 1 \%$, to the parity
   nonconservation (PNC) amplitude in heavy atoms, which makes them
   important for an analysis of modern experimental data. The present
   paper summarizes recent progress in calculations of the QED
   radiative corrections to PNC. Over the last few years the
   experimental data of the Boulder group for the $6s-7s$ PNC
   amplitude in $^{133}$Cs, which remains the most accurate in the area
   of atomic PNC, were presumed to be able to challenge the standard
   model.  The radiative corrections eliminate this possibility, their
   accurate determination brings the atomic experimental data for the
   mentioned transition in agreement with the standard model.  The
   paper presents also a brief description of the methods used for
   calculations of the QED corrections. Some of them, that have been
   recently designed for applications to the PNC problem,
   may be of interest in other areas where QED corrections are
   studied.
    
   Several atoms and several transitions have been studied
   experimentally in relation to PNC. As was mentioned, the most
   accurate experimental data was obtained for the PNC $6s-7s$ PNC
   transition in $^{133}$Cs.  Studies of this transition initiated by
   Bouchiat and Bouchiat \cite{bouchiat_74}, were continued and
   developed in Paris
   \cite{bouchiat_82,bouchiat_84,bouchiat_85,bouchiat_86,%
     bouchiat_86_1,bouchiat_02} and in Boulder
   \cite{boulder_85,boulder_86,boulder_88,wood_97}. The latest work of
   Wood {\it et al} \cite{wood_97} reduced the experimental error to
   $0.35\%$ that remains the best accuracy in atomic PNC experiments.
   These results inspired interest in very accurate atomic PNC
   calculations that, eventually, led to recognition of the fact that
   the QED corrections contribute at the necessary level of accuracy.
   Other experiments on PNC have been carried out for several
   transitions in several atoms. This includes the $6p_{1/2} -
   7p_{1/2}$ transition in $^{205}$Tl
   \cite{berkeley_79,berkeley_81,berkeley_85}, and the $6p_{1/2} -
   6p_{3/2}$ transition in $^{205}$Tl
   \cite{oxford_91_Tl,oxford_95,seattle_95}, the $^3P_0 -^3P_1$
   transition in $^{208}$Pb \cite{seattle_83,seattle_93,oxford_96} and
   two transitions in $^{209}$Bi, $^4S_{3/2} - ^2D_{5/2}$ measured in
   pioneering works of Barkov and Zolotorev
   \cite{novosibirsk_78,novosibirsk_78_1,novosibirsk_79,novosibirsk_80}
   and in \cite{moscow_84,oxford_87_5/2,oxford_93} as well as the
   $^4S_{3/2} - ^2D_{3/2}$ transition
   \cite{seattle_81,oxford_87_3/2,oxford_91_Bi}. The comprehensive
   theoretical background on PNC as well as a detailed account of
   events which led to a discovery of PNC in atoms (that was also the
   first observation of PNC in neutral currents) can be found in the
   book of Khriplovich \cite{khriplovich_91}.
   
   Errors of final quantities that can be extracted from experimental
   results on PNC in atoms crucially depend on the accuracy of atomic
   calculations.  The most difficult and cumbersome part of the
   theoretical research is the atomic structure calculations that take
   into account the many-electron nature of the atomic wave function
   in heavy atoms.  A foundation for the modern theoretical
   description of many-electron correlations in the PNC problem was
   laid out in the works \cite{dzuba_89,blundell_90} which also give
   references to previous papers. The calculations of the
   many-electron correlations in the $6s-7s$ transition in $^{133}$Cs
   in these works gave very close results, with the error in both sets
   of calculations estimated as $\sim 1 \% $.  The problem was
   revisited in several recent publications that supported the results
   of \cite{dzuba_89,blundell_90} improving the error.  Improvement of
   the theoretical error was initiated in Refs.
   \cite{bennett_wieman_99} that compared a number of experimentally
   measured quantities, such as polarizabilities, dipole amplitudes
   and hyperfine constants, with theoretical predictions of Refs.
   \cite{dzuba_89,blundell_90}, concluding that the theoretical error
   of these works was probably better than it had been anticipated, at
   the level of $0.4 \%$.  This conclusion allowed the authors of
   \cite{bennett_wieman_99} to reveal a deviation that existed between
   the experimental data of \cite{wood_97} and the standard model
   ($\sim 2 \sigma $).  The following theoretical works
   \cite{kozlov_01,dzuba_01,dzuba_02} demonstrated that indeed, the
   theoretical error that originates from the manyelectron
   correlations can be reduced down to $\sim 0.5 \% $ for the $6s-7s$
   transition in $^{133}$Cs, thus matching the experimental accuracy
   of Ref.  \cite{wood_97}.  The net result of the experimental and
   theoretical progress at this period of time was a pronounced
   deviation, of the order of $2 \sigma$, between predictions of the
   standard model and experimental data that was proved to be not
   related to uncertainties imposed by the complicated manyelectron
   nature of the problem.
    
   The progress outlined above indicated that either the standard
   model needed improvement, or some unrecognized, underestimated
   factors should come into play.  A careful analysis revealed that
   there was an overlooked factor well within the framework of the
   standard model, namely the conventional QED corrections to the PNC
   amplitude.  Obviously, they have always been kept in mind, but for
   a long period of time they were estimated as negligible in the PNC
   problem.  From the perspective of the present day situation the
   fact that they play an important role can be considered as a
   general statement.  Historically, however, its realization stemmed
   from several works devoted to different, separate phenomena that
   contribute to the QED corrections.  Derevianko \cite{derevianko_00}
   found that the Breit corrections to electron-electron interaction
   in heavy atoms are much larger than it had been anticipated, giving
   $-0.6 \% $ for the $6s-7s$ PNC amplitude in $^{133}$Cs.  This
   result was obtained numerically, and later confirmed in a number of
   papers \cite{dzuba_harabati_01,kozlov_01, sushkov_01} both
   numerically and analytically.  The account of the Breit corrections
   put the experimental data of \cite{wood_97} much closer to the
   standard model, but this was only the beginning of an intriguing
   zigzag road of research.  An analysis of the Breit corrections by
   Sushkov \cite{sushkov_01} revealed that one can expect the QED
   vacuum polarization to be as important as the Breit corrections.
   Johnson {\it et al} \cite{johnson_01} for the first time proved
   that indeed, the vacuum polarization is very important.  For the
   $6s-7s$ PNC amplitude in $^{133}$Cs atom they found that it gives
   $0.4 \%$, as confirmed in
   \cite{milstein_sushkov_01,dzuba_01,kf_jpb_02}. From a purely
   pragmatic, numerical point of view the vacuum polarization mostly
   compensated for the contribution of the Breit corrections, bringing
   the experimental data for the $^{133}$Cs atom back in contradiction
   with the standard model.
    
   The most difficult and, correspondingly, challenging part of the
   QED corrections to the PNC amplitude is presented by the QED
   self-energy corrections. We are using this term here in a sense
   that embraces the vertex corrections as well. One could have
   anticipated these corrections to be large from the very beginning.
   The self-energy corrections are known to give a large contribution
   to a number of phenomena in atomic physics, including the Lamb
   shift, the hyperfine interaction, and the energy shift due to the
   finite nuclear size (FNS). The latter phenomenon will play a
   particular role in the following discussion.  There is no physical
   reason indicating that the PNC amplitude should be any different.
   The sign of the self-energy corrections is also predetermined.  The
   QED self-energy is known to be mostly repulsive (this is definitely
   true for the $s_{1/2}$ state that should play a very important
   role) in contrast to the attractive vacuum polarization.  Therefore
   the self-energy corrections should make the electron wave function
   smaller at the nucleus and, consequently, give a negative
   contribution to the PNC amplitude.
    
    However, the aforementioned reasons were confronted by
    counter-arguments indicating that the self-energy corrections
    should be small. One reason, briefly mentioned in \cite{johnson_01},
    relies on the fact that in the plane-wave approximation
    the self-energy corrections are small, of the order of $0.1 \% $
    for $^{133}$Cs atom, as was found by Marciano and Sirlin
    \cite{marciano_sirlin_83} and Lynn and Sandars
    \cite{lynn_sandars_94}. One can argue that account of the
    nuclear Coulomb field should not produce any dramatic enhancement,
    and the self-energy corrections should, probably, remain small.
    The argument has its merit, it is known to be correct for the case
    of the vacuum polarization, where higher-order corrections are
    suppressed, see section \ref{vacuum and related}. However, the
    self-energy corrections behave differently. The well-studied case
    of the hyperfine interaction is an example. It is known from
    both analytical and numerical calculations, see Refs.
    \cite{blundell_97,sunnergren_98} and references therein, that in
    this problem the manifestations of the Coulomb field in the
    self-energy corrections in heavy atoms are more pronounced than a
    contribution of the plane-wave approximation, the latter one is
    given by the known $Z$-independent Schwinger term $\alpha/2\pi $,
    while the former rise quickly with the nuclear charge
    \footnote{This example for some time was part of a theoretical
    folklore. One of us (M.K.), who was a newcomer in the field of
    PNC, heard it first from M.G.Kozlov. At that time the validity of
    this argument could not probably be fully appreciated because
    another folklore wisdom claimed that the PNC should be different
    from everything studied previously due to a different operator
    structure, which is not the case, as is well established now, see
    Fig.\ref{four}.}.
    
  Another case, that looked even stronger, against a possible role of
  the self-energy corrections in the PNC problem was made by Milstein
  and Sushkov in Ref.\cite{milstein_sushkov_01}. They presented
  results of direct analytical calculations claiming that the
  self-energy corrections are positive and small, concluding that the
  experimental data on the $6s-7s$ PNC amplitude in the $^{133}$Cs
  atom are in contradiction with the standard model. The impact of
  this work was strong, for a long period of time it was considered as
  a reliable argument indicating that the experimental data can
  challenge the standard model. The opposite opinion expressed in Ref.
  \cite{dzuba_01}, which presented a model for rough estimation of the
  self-energy corrections, was not considered as convincing, probably
  because the proposed model neglected the vertex corrections and was
  not gauge invariant.
    
    The analytical calculations of Ref.\cite{milstein_sushkov_01} seemed
    to prove that the self-energy corrections are small and positive,
    while some physical arguments mentioned above indicated
    otherwise. To find the way out of this controversy we proposed in
    Ref.  \cite{kf_prl_02} an approach that avoided difficult
    analytical calculations.  This work derived a new equality, we
    will call it the chiral invariance identity due to reasons
    explained in section \ref{equality}.  It expressed the self-energy
    corrections for the PNC amplitude in terms of the self-energy
    corrections to the energy shifts due to FNS. The latter ones were
    calculated numerically in Refs.
    \cite{johnson_soff_85,blundell_92,%
      blundell_sapirstein_johnson_92,cheng_93,lindgren_93,cheng_02}.
    Thus the found equality allowed one for the first time to find the
    accurate self-energy corrections for the PNC amplitude for heavy
    atoms. As one could have anticipated, they were large and
    negative. In particular, for the $6s-7s$ transition in${133}$Cs we
    found $-0.73(20)\% $ that brought the experimental data of Wood
    {\em et al} in agreement with the standard model.  The uncertainty
    of this result was mainly due to errors of the numerical results
    of Ref.  \cite{cheng_93} that we used as input data. The most
    pronounced among them is the error for the self-energy corrections
    to the energy shift induced by the FNS in the $p_{1/2}$ wave.  The
    authors of \cite{cheng_93}, as well as others in the relevant
    papers above, estimated their results as very reliable for heavy
    atoms $Z\sim 90$, and less accurate for lighter atoms, making
    numerical calculations below $Z=55$ difficult.
    
    In order to verify and refine our results, there were performed
    analytical calculations of the leading, linear term in the $Z
    \alpha $-expansion for the self-energy (and vertex) corrections
    for the PNC amplitude \cite{k_jpb_02}. The result of this work
    confirmed that of \cite{kf_prl_02}, the self-energy corrections
    were found to be large and negative, in clear contradiction with
    Ref.  \cite{milstein_sushkov_01}. Combining the data of
    \cite{kf_prl_02}, which is accurate for heavy atoms, with the
    analytical data of \cite{k_jpb_02} that is reliable for light
    atoms, Ref.  \cite{k_jpb_02} found the self-energy corrections for
    the PNC amplitude for all atoms.  This analysis also allows one to
    reduce the impact of the above mentioned errors of the data
    extracted from \cite{cheng_93}.  The result for the $6s-7s$
    transition in $^{133}$Cs was $-0.9 \% $, in good agreement with
    $-0.73(20) \% $ of \cite{kf_prl_02}.  These two works, using
    different techniques, unambiguously demonstrated that the
    experimental data of Wood {\it et al} \cite{wood_97} support the
    standard model.
    
    The contradiction that existed between this statement and
    conclusions of Ref.  \cite{milstein_sushkov_01} was, fortunately,
    eradicated by the the work of Milstein {\it et al}
    \cite{milstein_sushkov_terekhov_02} that appeared shortly after
    \cite{k_jpb_02}.  Milstein {\it et al} arrived at results that are
    close to ours \cite{kf_prl_02,k_jpb_02}. For the $6s-7s$
    transition in $^{133}$Cs they found the self-energy corrections to
    be $-0.85\% $. A careful comparison of the results of the two
    groups presented in Ref.  \cite{kf_comm_02} revealed a similar
    agreement for all values of $Z$. In this latter work it was also
    proposed that the remaining small discrepancy is due to
    limitations of accuracy in calculations of Milstein {\it et al}
    that are based on $Z\alpha$-expansion ($Z\alpha = 0.4$ for Cs).
    This conclusion has recently been proved true by Milstein {\it et
      al} who refined their calculations in
    \cite{milstein_sushkov_terekhov_Log_all_orders}.  Now, for a vast
    area of the nuclear charges $Z$ the results of Milstein {\it et
      al} for the self-energy corrections agree with ours very
    closely, the remaining deviation is below $9 \% $.  The deviation
    manifests itself more prominently for large $Z$, and we believe
    that it is still due to those terms of the $Z \alpha $-expansion
    that remain unaccounted for by the analytical calculations of
    \cite{milstein_sushkov_terekhov_Log_all_orders}.  Thus, using
    several different techniques, the two teams arrived to one and the
    same result for the self-energy corrections. The fact that the
    results of Milstein {\it et al} converged to ours after a period
    of initial deviation makes the final results even more convincing.

    The plan of this paper is to discuss in some detail the radiative
    corrections, summarizing all available methods and data.  As an
    introduction to this study let us first present very briefly
    several known facts about relativistic corrections in general. The
    main relativistic correction is, obviously, the correction that is
    due to the relativistic nature of the Dirac equation that governs
    the electron propagation.  It is usually taken care of in the
    initial approximation that relies on the Dirac-Hartree-Fock
    approach, using then the machinery of the many-body theory to
    account for the electron correlations.  The next necessary step is
    to include the relativistic correction to the photon propagator
    that describes the electron-electron interaction. In the
    nonrelativistic approach this propagator is approximated by its
    Coulomb part. Relativistic effects can be taken care of
    perturbatively, using the Breit corrections for this propagator.
    This approximate approach is convenient because the Breit
    correction can also be incorporated in the initial Hartree-Fock
    mean field and then dealt with by conventional many-body methods.
    At the same time this approximation provides sufficient accuracy.

The next layer of corrections originates from the QED radiative
corrections.  Numerically, as well as parametrically, they may be
comparable with a contribution of the Breit corrections, but it is
convenient to consider them separately.  The radiative corrections in
the lowest order of the perturbation theory are
divided into two classes.  One of them constitutes the QED vacuum
polarization, another one consists of the self-energy and the vertex
corrections. These latter two corrections are often grouped together
and called the e-line corrections.

The discussion below focuses entirely on the QED radiative
corrections. This presumes that every other above mentioned
relativistic correction has been reliably taken care of in the
previous research on the atomic PNC problem, see the references discussed
above.  The present paper demonstrates that {\it all} radiative
corrections have recently been taken into account and calculated
reliably.  Thus, combining the radiative corrections with the previous
theoretical data one can be absolutely certain that {\it every}
possible relativistic correction in many-electron atoms is accounted
for. This done, the theory comes up with an important conclusion
\cite{kf_prl_02,k_jpb_02}: the most accurate available experimental
data for parity nonconservation of Wood {\it et al} \cite{wood_97}
fully support the standard model.



\section{ Vacuum polarization and related phenomena }
    \label{vacuum and related}
    
\subsection{Variation of the electron wave function due to short-range
    potentials}
   \label{wave function}
   
   The effective Hamiltonian $H_{\mathrm {PNC}}$ induced by the
   $Z$-boson exchange that describes the PNC part of the weak
   interaction between the atomic electrons and the nucleus can be
   presented in the following form (the relativistic units $\hbar = c
   =1,~e^2=\alpha$ are used, if not stated otherwise)

   \begin{equation}\label{05} H_{\mathrm {PNC}} = \frac{1}{ 2 \sqrt{2}
}\, G_{\mathrm F} \, Q_{\mathrm W}\, \rho(r)\,\gamma_5 ~.
   \end{equation} 
   Here $G_{\mathrm F}$ and $ Q_{\mathrm W} $ are the Fermi constant
   and the nuclear weak charge, and $\rho(r)$ is the nuclear density.
   The PNC amplitude describes the matrix element that mixes two
   states of opposite parity for a valence atomic electron. According
   to Eq.(\ref{05}) the electron wave function in this matrix element
   must be taken on the nucleus.  Therefore the largest PNC amplitude
   arises due to the mixing of the $s_{1/2}$ and $p_{1/2}$ partial
   waves $ \langle \psi_{n,s,1/2}| H_{\mathrm {PNC}} | \psi_{n',p,1/2}
   \rangle $, here $n,n'$ refer to the electron energy states, the
   electron wave functions are the Dirac spinors

   \begin{equation}\label{spinor}
     \psi_{nlj}({\bf r}) =
     \left(
       \begin{array}{cc} f_{nlj}(r)\Omega_{jl\mu}({\bf n}) \\
          i g_{nlj}(r)\Omega_{j\tilde l \mu}({\bf n})
       \end{array} \right)~,
   \end{equation}
   $f_{nlj}(r)$ and $g_{nlj}(r)$ are the large and small radial
   components of the spinor, $\Omega_{j l \mu}({\bf n})$ and
   $\Omega_{j\tilde l \mu}({\bf n})= -({\mbox {\boldmath
       $\sigma$}}\cdot {\bf n})$ $\Omega_{j l m}({\bf n})$ are the
   spherical spinors, and $l+\tilde l =2 j$.

   In applications the electron wave function is known in some
   reasonably good approximation (say, the Hartree-Fock-Dirac initial
   approximation plus contributions of the many-electron
   correlations). Inevitably there exist a number of perturbations
   that are not included in this approximation and can affect the
   electron wave function. The QED vacuum polarization is one of such
   perturbations, some others are considered below. In many cases the
   perturbation is localized in the vicinity of the nucleus, at
   distances $r$ that are much smaller than the Bohr radius of the
   K-shell $r \ll 1/(\alpha Z m)$. This is definitely the case for the
   QED vacuum polarization localized at $r \le 1/m$. For this range of
   distances the behavior of the electron wave function is quite simple
   because it is governed mostly by the pure Coulomb field of the
   nucleus. One can anticipate therefore that the influence of the
   perturbation on the electron wave function can be presented in
   simple terms, directly via the perturbative potential. This
   anticipation proves true, there exists a simple way to account for
   the perturbation that was suggested in Ref.\cite{kf_jpb_02}.
   
   Note that it suffices to find the relative variation of the
   electron wave function induced by a perturbation. Moreover, for
   small distances $r \ll 1/(\alpha Z m)$ the relative variation does
   not depend significantly on $r$. Therefore it is sufficient to find
   the relative variations of the wave function at the origin, i. e.
   to find $\delta f_{nlj}(0)/f_{nlj}(0)$ and $\delta
   g_{nlj}(0)/g_{nlj}(0)$, where $\delta f_{nlj}(r), \delta g_{nlj}(r)
   $ are the variations of the large and small components of the wave
   function due to the considered perturbation, and the relative
   variations at the origin are taken in the limit $r\rightarrow 0$. The
   limit is necessary because either the large or small components of
   the wave function vanish at the origin, see Eq.(\ref{bc}) below.
   Ref.  \cite{kf_jpb_02} found for the relative variations of the
   wave functions the following result

\begin{equation}\label{final}
\frac{\delta f(0)}{f(0)}=\frac{\delta g(0)}{g(0)}=
-\frac{m}{\hbar^2}\,\int_0^\infty V(r)\,(a+kr)\,{\rm d}r~.
   \end{equation}
   Here and below the indices $nlj$ for the wave functions are
   suppressed to simplify notation, $a$ is a parameter with length
   dimension and $k$ is a dimensionless coefficient

\begin{eqnarray} \label{a}
a &=&  \frac{Z\alpha}{\gamma}\,\frac{\hbar}{mc}~,
\\ \label{k}
k &=&  \frac{2\kappa(2\kappa-1)}{\gamma(4\gamma^2-1)}~,
   \end{eqnarray}
   $V(r)$ is the perturbative potential.  Parameters $\gamma, \kappa$
   are defined conventionally $\gamma = (1-(Z \alpha )^2)^{1/2}$ and
   $\kappa = l(l+1)-j(j+1)-1/4 = \pm (j+1/2)$.  Relations
   (\ref{final}),(\ref{a}) are presented in absolute units, to make
   them more convenient for different applications below.  The simple,
   clear formula (\ref{final}) solves the problem formulated above,
   presenting a variation of the wave function at the origin in
   transparent terms, over the two first moments of the potential.  It
   is instructive to consider Eq.(\ref{final}) in the nonrelativistic
   limit that reads

 \begin{equation}\label{nonrelat}
 \frac{ \delta \psi( 0 ) }{ \psi(0) } =
-\frac{m}{\hbar^2} \,\frac{2}{2l+1}\int_0^\infty V(r)\,r\,{\rm d}r~.
    \end{equation}
    Here $\psi(r)$ is the Schroedinger nonrelativistic wave function.
    Deriving Eq.(\ref{nonrelat}) from (\ref{final}) one uses the fact
    that according to (\ref{a}) the parameter $a$ goes to zero in the
    limit $Z\alpha \rightarrow 0$, while from (\ref{k}) one finds $k
    \rightarrow 2/(2l+1)$.  There is a simple short-cut derivation
    that leads to (\ref{nonrelat}). One expresses the variation of the
    Schroedinger electron wave function via the Green function $G$

\begin{equation}\label{delta}
\delta \psi(0) = \int G({\bf 0},{\bf r}')\,V(r')\, 
\psi({\bf r}')\,{\rm d} ^3 r'~,
    \end{equation}
    approximating the Green function by the Green function for free
    motion $G^{(0)}$

 \begin{eqnarray}\label{G0} G({\bf r},{\bf r}') &\rightarrow & G^{(0)}_l
 ({\bf r},{\bf r}') \simeq
\\ \nonumber
&-&\frac{2m}{\hbar^2} \frac{1}{2l+1}
 \frac{r_<^l}{r_>^{l+1}} \sum_{m=-l}^{l} Y_{lm}({\bf n})
Y_{lm}^\dag({\bf
 n}') ~, 
   \end{eqnarray} 
   where $G^{(0)}_l({\bf r},{\bf r}')$ is the Green function for the
   free motion in the $l$-th partial wave, ${\bf n} = {\bf r}/r,~ {\bf
     n}' = {\bf r}'/r'$, and the last identity in Eq.(\ref{G0}) takes
   into account the fact that the binding energy of the valence
   electron is negligible for short distances.  Remembering that for
   nonrelativistic motion the wave function behaves like
   $\psi_{l}({\bf r}) \propto Y_{lm}({\bf n})\,r^l,~r \rightarrow 0$
   one immediately derives Eq.(\ref{nonrelat}) from Eq.(\ref{delta}).
    
   The point to be noted in this simple derivation is the dominance of
   the kinetic term in the nonrelativistic limit. For small distances
   the nonrelativistic kinetic energy behaves as $\propto 1/r^2$,
   being larger than the potential energy.  That is why Eq.(\ref{G0})
   holds in the nonrelativistic limit. In the relativistic approach
   the situation is different. The kinetic term behaves as $\propto
   1/r$, while the potential behaves as $Z \alpha /r$.  One has to
   expect henceforth that the contribution of the Coulomb potential
   energy to the final answer is to be suppressed compared with the
   contribution of the kinetic term only by a factor $Z \alpha $.
   This is exactly what happens in Eq.(\ref{final}). The second term
   $\propto kr$ on the right-hand side can be considered as a mere
   modification of the nonrelativistic result, while the first one
   $\propto a$ is the expected new term proportional to $Z \alpha $.
   Thus the structure of Eq.(\ref{final}) is predetermined.  The
   calculations of Ref.  \cite{kf_jpb_02} establish the coefficients
   in this relation as well as the dependence on higher powers of the
   $Z \alpha $-expansion that appear through the parameter $\gamma$.
    
    The nonrelativistic Eq.(\ref{nonrelat}) shows that the parameter
    that governs the variation of the wave function is $\int m
    V(r)r{\rm d}r$ that is almost an obvious result valid for a
    variety of quantum mechanical problems \cite{LLIII}. It is
    interesting that relativistic Eq.(\ref{final}) shows that there
    exists another relativistic parameter $\int ma V(r){\rm d}r$.  It
    is suppressed compared with the nonrelativistic one only by a
    factor $Z\alpha$ that is not small for heavy atoms. This
    suppression can be well compensated for if the potential increases
    at small distances which makes $ \int ma V(r){\rm d}r$ larger than
    $m \int V(r)r{\rm d}r$. In this case the new relativistic
    parameter becomes dominant, as the examples below demonstrate. The
    singular rise of the potential at small distances is typically
    smeared out by the finite nuclear size. Dealing with such singular
    potentials it is important to keep in mind that the first term on
    the right-hand side of Eq.(\ref{final}) is proportional to $Z$.
    Therefore it originates from those distances where the full
    strength of the electric field created by the nuclear charge is
    present, i. e.  from the region outside the nuclear core $r \ge
    r_{\mathrm N}$.  Inside the nucleus the Coulomb field created by
    the nuclear charge diminishes, making this region less important.
    In order to estimate its contribution, one can introduce for
    distances $r \le r_{\mathrm N}$ an effective nuclear charge $Z
    \rightarrow Z_\mathrm{ eff}= (r/r_\mathrm{N} )^3 \,Z$ that
      diminishes with the nuclear radius taking into account the
      decline of the electric field inside the nucleus.
      Correspondingly modifying the parameter $a$ in Eq.(\ref{final})

    \begin{equation}
      \label{Zeff}
      a \rightarrow a_\mathrm{    eff}= \left( 
     \frac{r}{ r_\mathrm{N} }\right)^3 \,a~,
    \end{equation}
    one can apply this equation for the region $r\le r_{\rm N}$. We
    will see that this approximate procedure gives sensible results
    for the vacuum polarization, see Eqs.(\ref{w})-(\ref{w2}).
    
    Let us apply Eq.(\ref{final}) to the PNC problem, in which one
    needs to calculate the PNC matrix element $ \langle
    \psi_{n,s,1/2}| H_{\mathrm {PNC}} | \psi_{n',p,1/2} \rangle$.  Both
    $s_{1/2}$ and $p_{1/2}$ wave functions are influenced by the
    perturbation $V(r)$. Therefore the relative variation of the
    PNC amplitude can be found simply by adding variations of
    $s_{1/2}$ and $p_{1/2}$ states given in Eq.(\ref{final})

 \begin{eqnarray}\label{weak}
\delta_{\mathrm {PNC},\,V} &\equiv & \frac{\delta
 \langle \psi_{s,\,1/2}|H_{\mathrm {PNC}} |\psi_{p,\,1/2}\rangle} 
{\langle \psi_{s,\,1/2}|H_{\mathrm {PNC}} |\psi_{p,\,1/2}\rangle} =
\\ \nonumber
&&-\frac{2m}{\hbar^2}\,\int_0^\infty V(r)\,\left(a+\frac{2}{3}
 kr\right)\,{\rm d}r~.
    \end{eqnarray}
    Here $\delta \langle \psi_{s,\,1/2}|H_{\mathrm {PNC}}
    |\psi_{p,\,1/2}\rangle$ is the variation of the PNC matrix element
    due to variations of the wave functions created by the potential
    $V(r)$.  Deriving this result we take into account that essential
    parameters for the $s_{1/2}$ and $p_{1/2}$ states are
    $\kappa_{s}=-1,~\kappa_{p}=1, ~\gamma_s = \gamma_p =
    (1-(Z\alpha)^2)^{1/2}\equiv \gamma$ and define $k$ in (\ref{weak})
    to be $k\equiv k_s= 6/\Big( \gamma(4\gamma^2-1) \Big)$.
    
    It was assumed in Eq.(\ref{final}) that the perturbative potential
    is a scalar. In applications it is necessary also to deal
    with more sophisticated, tensor-type perturbations. The electric
    and magnetic fields created by the nuclear dipole, quadrupole and
    higher moments give examples of such perturbations. It is
    interesting that Eq.(\ref{final}) can be generalized to describe
    the tensor-type perturbations in simple clear terms.  Suppose that
    we have the valence electron described by the Dirac spinor
    $\psi_{lj}({\bf r})$ (the energy of the valence electron is so low
    that we can neglect it, suppressing the irrelevant index $n$).
    Suppose that there is some tensor perturbation $V({\bf r})$.  When
    we consider the PNC amplitude we need to evaluate the wave
    function at the nucleus. There are only two partial waves that
    remain large at the origin, $s_{1/2}$ and $p_{1/2}$. This means
    that in  the multipole expansion of the $V({\bf
    r})\psi_{lj}({\bf r})$ term in the Dirac equation only these
    two partial waves contribute. The angular structure of these
    partial waves is predetermined, compare Eq.(\ref{spinor}),
    therefore we may write
    
    \begin{equation}
      \label{partial}
    [V({\bf r})\psi_{lj}({\bf r})]_{L,1/2} =
     \left(
       \begin{array}{cc} F_{L,1/2}(r)\Omega_{1/2,L,\mu}({\bf n}) \\
          i \,G_{L,1/2}(r)\Omega_{1/2,\tilde L, \mu}({\bf n})
       \end{array} \right)~,
    \end{equation}
    where $[V({\bf r})\psi_{lj}({\bf r})]_{L,1/2}$ means the
    admixture of the $L_{1/2}$ state in the term $V({\bf
      r})\psi_{lj}({\bf r})$.  Eq.(\ref{partial}) shows that the
    contribution of the term $V({\bf r})\psi_{lj}({\bf r})$ to the
    Dirac equation is expressed in terms of the scalar functions
    $F_{L,1/2}(r)$ and $G_{L,1/2}(r)$ with $L=0,1$. We can
    introduce now some {\it effective} potentials $V^{\mathrm
      {eff}}_L(r)$, $W^{\mathrm {eff}}_L(r) $ that satisfy the
    following conditions

   \begin{eqnarray} \label{effect1} F_{L,1/2}(r) &=& V^{\mathrm
     {eff}}_L(r) f_{L,1/2}(r)~, \\ \label{effect2} G_{L,1/2}(r) &=&
     W^{\mathrm {eff}}_L(r) g_{L,1/2}(r)~.  
     \end{eqnarray} 
     Here $f_{L,1/2}(r),~ g_{L,1/2}(r)$ are the known large and small
     components of the nonperturbed wave function for the valence
     electron in the $L_{1/2}$ partial wave. The procedure described
     shows that the term $V({\bf r})\psi_{lj}({\bf r})$ in the Dirac
     equation at small distances can be described by the effective
     potentials $V^{\mathrm {eff}}_L(r)$ and $ W^{\mathrm {eff}}_L(r)$
     (which can be different for the large and the small components
     for an arbitrary perturbation).  Correspondingly, the variation
     of the wave function $\delta \psi_{lj}({\bf r})$ at small
     distances can be expressed via its multipole expansion in which
     only the $s_{1/2}$ and $p_{1/2}$ states are retained

    \begin{equation}
      \label{wf,partial}
    \delta \psi_{lj}({\bf r}) \simeq
     \sum_{L=0,1}
      \left(
       \begin{array}{cc} \delta f_{lj;\, L,1/2}(r)
                       \Omega_{1/2,L,\mu}({\bf n}) \\
          i \delta g_{lj;\, L,1/2}(r)\Omega_{1/2,\tilde L, \mu}({\bf n})
       \end{array} \right)~.
    \end{equation}
    Here the functions $\delta f_{lj;\, L,1/2},~\delta g_{lj;\,
      L,1/2}$ describe the admixture of the $L_{1/2}$ states to the
    function $\psi_{lj}({\bf r})$ that arise due to the tensor-type
    perturbation.  They are to be found from the Dirac equation. We
    see that the complicated multipole problem is reduced to the
    problem with the scalar potential, the only distinction is that
    the role of the scalar potential plays the effective potentials
    $V^{\mathrm {eff}}_L(r),~W^{\mathrm {eff}}_L(r)$ defined above.
    However, this distinction makes no difference, we introduce the
    effective scalar potentials in the Dirac equation and solve it
    regardless of the nature of the potentials. We can  rely
    on the discussion in \cite{kf_jpb_02} to derive formulas
    similar to Eq.(\ref{final}).  The result reads

 \begin{eqnarray}\label{eff1}
\frac{\delta f_{lj;\, L,1/2}(0)}{f_{L,1/2}(0)}&=&
-\frac{m}{\hbar^2}\int_0^\infty 
\!\!\!\!V^{\mathrm {eff}}_{\mathrm L}(r)
(a+kr){\rm d}r, 
\\ \label{eff2}
\frac{\delta g_{lj;\, L,1/2}(0)}{g_{L,1/2}(0)} &=&
-\frac{m}{\hbar^2}\int_0^\infty 
\!\!\!\!W^{\mathrm {eff}}_{\mathrm L}(r)
(a+kr){\rm d}r.
    \end{eqnarray}
    Remember that here $\delta f_{lj;\, L,1/2}(0),~\delta g_{lj;\,
      L,1/2}(0)$ describe the admixture of $L_{1/2},~L=0,1$ states
    that arise in the wave function $\psi_{lj}({\bf r})$ of the
    valence electron in the state with the quantum numbers $lj$ under
    the influence of the tensor-type perturbation $V({\bf r})$, while
    $f_{L,1/2}(r),~g_{L,1/2}(r)$ describe the wave function of the
    valence electron in the state with the quantum numbers $L_{1/2}$.
    The latter ones are the same as the ones used in
    Eqs.(\ref{effect1}),(\ref{effect2}) to define the effective
    potentials. This makes our results presented in
    Eqs.(\ref{eff1}),(\ref{eff2}) independent of the normalization of
    the wave functions $f_{L,1/2}(r),~g_{L,1/2}(r)$. In
    particular, they are independent of the energy of the valence
    electron (as they should be since this energy is too low to produce
    any significant effects for small distances).  The quantum numbers
    in Eqs.(\ref{a}),(\ref{k}) that define the parameters $a,k$ in the
    right-hand sides of Eqs.(\ref{eff1}),(\ref{eff2}) correspond to
    the state $L_{1/2},~L=0,1$. We see that for an arbitrary
    tensor-type potential the result remains very simple. The only
    modification, compared with the simplest scalar case, is that one
    needs first to define the effective potentials using Eqs.
    (\ref{effect1}),(\ref{effect2}).

    Summarizing, we described in this section simple useful
    procedures that express the variation of the valence electron wave
    function at the origin directly in terms of the perturbative
    potential.

\subsection{QED vacuum polarization}
   \label{vacuum}
   
   The vacuum polarization in the lowest order of  perturbation
   theory is described by the Uehling potential $V_{\mathrm VP}$
   \cite{uehling}

\begin{eqnarray}\label{uehling}
V_{\rm VP}(r) &=&
-\frac{2\alpha}{3\pi} \,\left(\frac{Ze^2}{r}\right)\, 
\\ \nonumber
&\times& \int_1^\infty e^{-2mr\zeta}Y(\zeta)\,{\rm d}\zeta, \\ \label{Y}
 Y(\zeta) &=& \left(1+\frac{1}{2\zeta^2}\right)
\frac{ \sqrt{ \zeta^2-1}}{\zeta^2}.
    \end{eqnarray}
    Importantly, the Uehling potential (\ref{uehling}) is singular at
    the origin, its asymptotic for  $mr \ll 1$ reads 

\begin{equation}
\label{uo}
V_{\mathrm {VP}}(r) = - \frac{2 \alpha}{3\pi} \, \left( \frac{Ze^2}{r}
\right) \, \left( \ln\frac{1}{mr}-C-\frac{5}{6}\right), 
   \end{equation} the singularity is smeared out by the nuclear finite
   size $r_{\mathrm N}$.  In Eq.(\ref{uo}) $C = 0.577\dots$ is the
   Euler constant.  The function $\ln mr $ arises due to conventional
   scaling of the QED coupling constant $e^2$ that manifests itself at
   short distances. Alternatively, this fact is referred to as the
   unscreening of the nuclear charge at small distances.  For large
   distances $mr \gg 1~$ the Uehling potential exponentially decreases

\begin{equation}
   \label{exp} V_{\rm VP}(r) = - \left( \frac{Ze^2}{r} \right) \,
\frac{\alpha \, \exp(-2mr) }{ 4 \pi^{1/2} \, (mr)^{3/2} }~. 
   \end{equation} 
   In order to find the influence of the vacuum polarization on the PNC
   amplitude we substitute Eq.(\ref{uehling}) into Eq.(\ref{weak}).
   Integration over the variable $r$ is performed analytically
   resulting in the following expression for the relative correction
   to the PNC amplitude

\begin{eqnarray}\label{w}
\delta^{(\mathrm {VP}) } &=&
\delta^{( \mathrm{VP},\,1) } +
\delta^{(\mathrm{VP},\,2)},
   \end{eqnarray}
   where

\begin{eqnarray}
\label{w1}
\delta^{ ( \mathrm {VP},\,1 ) } &=&
\frac{Z \alpha^2}{\gamma} \left(
\,\frac{3}{4}\,\,\frac{1}{4\gamma^2-1} + \right.
\\ \nonumber
&&
\left. 
Z \alpha \,\frac{4}{3\pi}
\int_1^\infty E_1(2 mr_N \zeta) Y(\zeta) \,{\mathrm d}\zeta
\right)~,
     \\ \label{w2}
\delta^{(     \mathrm {VP,\,2) }     } &=&
- \frac{ Z \alpha }{ 2\gamma } \, 
( \,V_{ \mathrm{VP} } ( r_{\mathrm N} ) \, r_{ \mathrm N }\,)~.
     \end{eqnarray}
   Here the function $E_1(x)$ is defined conventionally
\begin{equation}
     \label{E1} E_1(x) = \int_1^\infty \exp(-xt)\,\frac{dt}{t}~.
    \end{equation} 
    Eq.(\ref{w1}), which was first found in \cite{kf_jpb_02}, gives
    the main contribution \footnote { Eqs.(43),(45) of
      \cite{kf_jpb_02} missed a factor of 2 in the argument of the
      $E_1$ function.  This resulted in an overestimation of the role
      of the vacuum polarization, that was claimed to be $0.47 \% $
      for the $^{133}$Cs atom, while the present paper predicts $0.40
      \% $.}.  Deriving it one restricts the integration of
    the singular term $ \int m a V_\mathrm{VP}$ in Eq.(\ref{weak}) to
    the region outside the nucleus $r\ge r_\mathrm{N}$. The large
    Coulomb field of the nucleus makes this region important.  Inside
    the nucleus $r\le r_\mathrm{N}$ the electric Coulomb field
    diminishes, suppressing the contribution of this region. In order
    to estimate the contribution of small distances $r\le
    r_\mathrm{N}$ we use a modification of Eq.(\ref{weak}) introducing
    into Eq.(\ref{final}) the effective parameter $a$ defined in
    Eq.(\ref{Zeff}).  The resulting integral in Eq.(\ref{final}) is
    saturated in the nuclear interior region in the vicinity of the
    nuclear surface, where the electric field remains large.  This
    fact allows one to take the values of $\gamma$ and
    $V_\mathrm{VP}(r)$ directly at the nuclear surface.  The described
    procedure results in estimate Eq.(\ref{w2}).
     
    Eqs.(\ref{w})-(\ref{w2}) give the weak interaction matrix element
    for an arbitrary atom in a transparent analytical form without
    fitting parameters.  Numerical results are easily obtained by a
    straightforward one-dimensional integration in Eq.(\ref{w1}).  One
    only needs to specify the nuclear size that can be taken
    conventionally as $r_N = 1.1  A^{1/3}$ fm, where $A$ is the
    atomic number \footnote{The coefficient in the relation $r_N =
      \eta \cdot A^{1/3}$ fm is usually taken in the region $\eta =
      1.1 - 1.2$.  Fixing its value at $\eta =1.1$ one makes a choice,
      though the variation in the mentioned region produces a small
      effect.}.  Alternatively, the right-hand side of Eq.(\ref{w}) can
    be calculated using an expansion in powers of $mr_N \ll 1$ that
    reads \cite{kf_jpb_02}

   \begin{eqnarray}
     \label{log2}
\!\!\!\!&&\delta^{( \mathrm {VP} ) } = 
\frac{\alpha}{\gamma} \left\{
   \,\frac{3}{4(4\gamma^2-1)}\,Z\alpha
   + \right.
\\ 
\nonumber
\!\!\!\!&&
\left.
\frac{2}{3\pi}(Z\alpha)^2\left[
   \left(\ln\frac{1}{mr_N}-C-\frac{5}{6}\right)^2 +
   0.759\right]  \right\},
   \end{eqnarray}
   where the omitted terms are $O(mr_{\mathrm N})$ \footnote{This
     expansion brings (\ref{w}) to a form that is close, but not
     identical to the one derived in \cite{milstein_sushkov_01}. We do
     not pursue the origin for this discrepancy since calculations in
     the cited paper were restricted by the logarithmic accuracy.}.
   The error of simplification of Eq.(\ref{weak}) to (\ref{log2})
   remains $ \le 7 \% $ for heaviest atoms, decreasing for lighter
   atoms. A very interesting feature revealed by Eq.(\ref{log2}) is
   the second power of the logarithmic factor in the term $\sim
   Z^2 \alpha^3  \ln^2(m r_{\mathrm N})$. This strong, double-logarithm
   enhancement was first found by Milstein and Sushkov
   \cite{milstein_sushkov_01}.  The derivation presented above makes
   the origin of this effect clear. The first power of the logarithmic
   function $\ln m r$ originates from the unscreening of the nuclear
   charge at small separations (\ref{uo}), while the second one
   appears from the relativistic parameter $ \int V(r)
   dr\propto \int dr \ln (mr)/r\propto \ln ^2 1/m r_\mathrm{N} $ in
   Eq.(\ref{weak}).

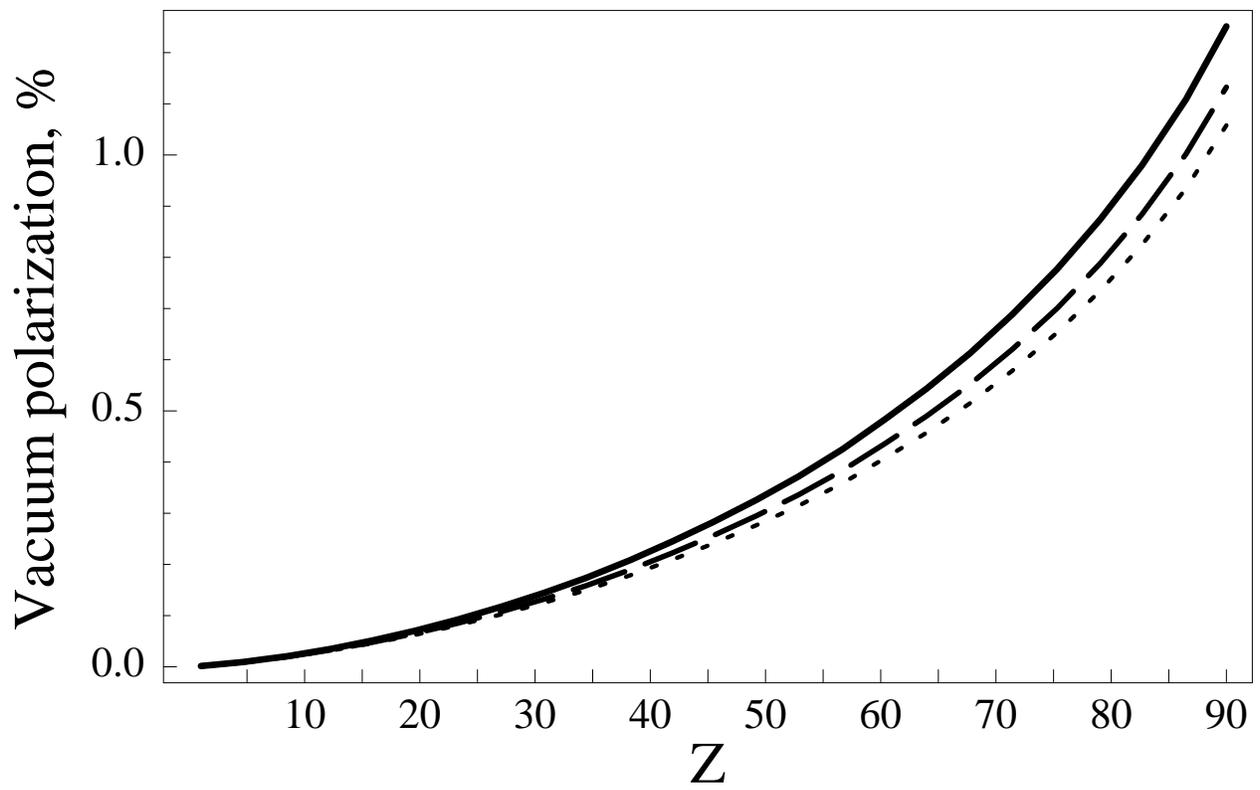
  \begin{figure}[t]
  \caption{ \label{one} 
    Relative contribution (\%) of the QED vacuum polarization to the
    PNC amplitude versus the nuclear charge.  Thick line - total
    correction predicted by Eq.(\ref{w}), dashed line - prediction of
    Eq. (\ref{w1}) that takes into account only events outside the
    nucleus, dotted line - results based on the expansion
    Eq.(\ref{log2}).  
}\end{figure}

   In Fig. \ref{one} we present the relative corrections due to the QED
   vacuum polarization $\delta^{(\mathrm{VP})}$ to the PNC matrix
   element calculated with the help of Eqs.(\ref{w})-(\ref{w2}). The
   corrections are positive and rapidly grow for heavy atoms. For the
   most interesting case of the $^{133}$Cs atom Eq.(\ref{w}) gives
   $\delta^{(\mathrm{VP})} =0.40 \% $.  For $^{205}$Tl, $^{208}$Pb,
   and $^{209}$Bi Eq.(\ref{w}) predicts corrections
   $\delta_{(\mathrm{VP})} = 0.93,~0.96,~0.99\% $ respectively.  For
   all four atoms the nuclear interior region gives 10 \% of the total
   correction, $\delta^{(\mathrm{VP},\,2)} = 0.1
   \delta^{(\mathrm{VP})} $, which means that to make results accurate
   numerically this region needs to be taken care of.

%
 
   Compare these results with other results reported recently.
   Johnson, Bednyakov and Soff \cite{johnson_01} calculated for the
   first time the correction due to the vacuum polarization for the
   $6s-7s$ PNC amplitude in $^{133}$Cs and found it to be $0.4\% $.  The
   result of \cite{johnson_01} includes, along with the variation of
   the weak matrix element, variations of the dipole matrix element
   and the corresponding energy denominators that, combined together,
   describe the $s_{1/2}-s_{1/2}$ amplitude measured experimentally.
   Dzuba {\it et al} \cite{dzuba_01} confirmed this result,
   calculating the correction $0.41 \% $, and supplied more details
   providing separate variations for all three quantities mentioned
   above. They found that variations of the dipole matrix elements and
   the energy denominators induced by the Lamb shift corrections,
   being not small, compensate each other almost completely. Thus the
   variation of the PNC amplitude proves to be equal to the variation of
   the PNC matrix element.  The numerical result for thallium found in
   \cite{df_03} gives a correction $0.94 \% $, to be compared with
   $0.93\% $ mentioned above.  Calculations of Milstein {\em et
     al} in Ref.\cite{milstein_sushkov_01} were restricted by the
   logarithmic accuracy that was further improved by using some
   constant as a fitting parameter to reproduce $0.4\% $ for
   $^{133}$Cs, in line with \cite{johnson_01}. For Tl
   Ref.\cite{milstein_sushkov_01} finds $0.9 \% $. This is slightly
   less than the above mentioned prediction of Eq.(\ref{w}) ($0.93 \%
   $). This (insignificant) discrepancy probably arises
   because the procedure adopted in \cite{milstein_sushkov_01}  does
   not account for the increase of the contribution of the nuclear
   interior region with the nuclear charge. We conclude that all
   reported calculations for the vacuum polarization correction are in
   very good agreement, distinctions between results derived by
   different methods and by different groups are all below $5 \%$.

   Let us discuss now possible corrections to the results discussed
   above. One of them originates from a modification of the vacuum
   polarization due to the finite nuclear size.  The corresponding
   generalization of the Uehling potential was given by Fullerton
   and Rinker \cite{fullerton_rinker}. In order to estimate its
   contribution we need to remember that the main $\propto \ln^2
   mr_\mathrm{N}$ term in Eq.(\ref{log2}) originates from the
   integration well outside of the nuclear radius where the Uehling
   potential is very close to the one of Fullerton and Rinker.
   Therefore this main part of the correction is not changed.
   Modifications could arise only for the integration in the vicinity
   of the nuclear surface, but this region by itself gives only a
   fraction ($\sim 10 \% $) of the total correction, while the
   relative difference of the Uehling potential and the Fullerton and
   Rinker potential on the nuclear surface is small. For all atoms it
   remains  $\le 5 \% $, as we found by direct numerical
   comparison of the Uehling potential with the Fullerton and Rinker
   potential. We conclude that the modification of the vacuum
   polarization due to the finite nuclear size is irrelevant.

     Another possible modification can originate from the QED
     higher-order corrections that contribute to the vacuum
     polarization. One of them stems from the fact that the Landau
     pole requires that the linear in $\ln mr$ term in
     Eq.(\ref{uo}) be followed by higher-order logarithmic
     terms.  However, the corresponding series runs in powers of $Z
     \alpha(\alpha \ln mr)^n, ~n=1,2 \dots$, giving  
     small contributions for higher order terms $n>1$.

     Consider now a modification of the Uehling potential that is due to
     the fact that the Coulomb field of the nucleus disturbs the virtual
     electron-positron pair.  This effect is conveniently described with
     the help of the Wichmann-Kroll potential \cite{wichmann-kroll}
     that, compared to Eq.(\ref{uehling}), describes the next term in
     the $Z \alpha $-expansion. For estimates it is sufficient
     to take the Wichmann-Kroll potential at small distances, where its
     asymptotic $mr\ll 1$ reads \cite{milstein_strakhovenko_83}:

  \begin{equation}\label{wk}
  V_{\rm WK}(r) = 0.092 \,\frac{2 \alpha}{3\pi}\,(Z \alpha)^2 \, \left
   ( \frac{Ze^2}{r} \right)~.  \end{equation} Using
   Eq.(\ref{weak}) one immediately finds from Eq.(\ref{wk}) the ratio of
   the correction induced by the Wichmann-Kroll potential to the Uehling
   correction

  \begin{equation}\label{ratio}
  \frac{ \delta ^ \mathrm{ ( VP ) } _ \mathrm {WK} } 
    { \delta^\mathrm{  (VP)  }_\mathrm{ U } } = 
  -0.184 \, 
  \frac{ ( Z \alpha )^2 }  {  \ln  \Big( 1/(mr_\mathrm{N} ) \Big)   }~,
     \end{equation}
     as was found in \cite{kf_jpb_02}.  For $^{133}$Cs this ratio is
     about $-0.007 $, demonstrating that the correction due to the
     Wichmann-Kroll potential is very small. Thus all higher order QED
     terms in the vacuum polarization are negligible.

     Alongside the vacuum polarization there exists also another
     polarization effect, namely the polarization of the nucleus.  One
     can estimate its influence on the electron wave functions and
     the PNC matrix element with the help of the conventional
     polarization potential $V_\mathrm{NP}$ that is created by the
     nucleus

     \begin{equation}
       \label{nucPol}
       V_\mathrm{NP} (r) = - \frac{ e^2\,  \alpha_\mathrm{d} }
     { 2 r^4   }~.
     \end{equation}
     Here $\alpha_\mathrm{ d}$ is the static nuclear dipole
     polarizability. It can be estimated roughly assuming that the total
     oscillator strength for the nuclear dipole excitations is located
     at the energy of nuclear giant dipole resonance
     $\Omega_\mathrm{d}$.  Then, using the conventional sum rule for the
     oscillator strengths one finds

     \begin{equation}
       \label{DipRes}
       \alpha_\mathrm{d} \simeq  \frac{ Z e^2 }
                                {  m_p \, \Omega_\mathrm{d}^2 }~,
     \end{equation}
     where $m_p$ is the proton mass.  Since the potential (\ref{nucPol})
     increases at small separations the main contribution in
     Eq.(\ref{weak}) is given by the first term in Eq.(\ref{weak}) which
     should be integrated outside the nuclear radius $r\ge r_\mathrm { N
     }$, as was explained above. Having this in mind, one substitutes
     (\ref{nucPol}) in (\ref{weak}) finding  an estimate for the
     influence of the nuclear polarization on the PNC matrix element

     \begin{equation}
       \label{PNCnucPol}
         \delta^\mathrm{ (NP) } \simeq 
  \frac{2}{3}\, \frac{Z^2 \alpha^3}
  {\gamma \,m_p\,\Omega_  \mathrm { d }^2 \,r_\mathrm{N}^3 } \,~.
     \end{equation}
     For the $^{133}$Cs atom we take $\Omega_\mathrm {d} \simeq 15$ MeV
     obtaining $\delta^{(\mathrm{ NP })} \simeq 0.01 \% $ that shows that
     the nuclear polarization is negligible. This conclusion agrees with
     estimates of Milstein and Sushkov \cite{milstein_sushkov_01}, who
     proposed that the effect is of the order of $\sim 0.1
     Z^{2/3}\alpha^2$ \footnote{ Eq.(12) of \cite{milstein_sushkov_01}
       suggests that the corrections are negative, which is probably a
       typo since the static polarization inevitably results in an
       increase of the effect.},  which numerically is close
     to Eq.(\ref{PNCnucPol}).

     Summarizing, we demonstrated above that the QED vacuum
     polarization gives a large positive correction ($\sim 1 \% $) to
     the PNC amplitude. Contributions of higher order QED corrections
     are much smaller. This means that the influence of the nuclear
     Coulomb potential on the virtual electron-positron pair is not
     pronounced, possibly due to opposite signs of the virtual
     particles. We will see below that for a single electron the
     situation is different, the Coulomb field does play an important
     role for the self-energy correction, see Eq.(\ref{197}).  The
     influence of the finite nuclear size on the vacuum polarization
     was found to be nonessential, the polarization can be well
     described by the simplest Uehling potential. Also negligible is
     the polarization of the nucleus.  We presented above in some
     detail analytical methods of calculation, paying less attention
     to purely numerical methods that have played a very important role
     in the problem, see \cite{johnson_01,dzuba_01,df_03}. Omissions
     of details in our presentation of numerical methods is justified
     by the fact that the vacuum polarization is described by a
     sufficiently simple potential, which can be incorporated in the
     initial Hartree-Fock-Dirac approximation for the atomic
     electrons, and then dealt with by conventional methods of 
     many-body theory. Thus the numerical approach follows
     the mainstream of atomic structure calculations discussed in some
     detail elsewhere, see for example \cite{dzuba_89,blundell_90}.
     Overall, the polarization is a well understood, robust
     effect that is well described by Eqs.(\ref{w})-(\ref{w2}).

  \section{ Self-energy and vertex corrections }

  \subsection{Chiral Invariance and QED corrections }
      \label{equality}
      Let us show, following Ref.\cite{kf_prl_02}, that there exists an
      interesting and useful relation that gives the QED radiative
      corrections to the PNC matrix element via similar radiative
      corrections to another quantity, the energy shifts of the atomic
      electron induced by the finite nuclear size (FNS). This relation can
      be written as

  \begin{equation}\label{ddd}
        \delta_{\mathrm {PNC},\,sp} = \frac{1}{2}\,  
        ( \delta_{ \mathrm {FNS},\,s } +
        \delta_{\mathrm {FNS},\,p })~,
       \end{equation}
       where $\delta_{ \mathrm {PNC},\, sp }$ is the relative radiative
       correction to the PNC matrix element between the $s_{1/2}$ and
       $p_{1/2}$ orbitals

  \begin{equation}\label{d} \delta_{\mathrm {PNC},\,sp} =
        \frac{\langle \psi_{s,\,1/2} | H_{\mathrm {PNC}}
        | \psi_{p,\,1/2}\rangle^ {{\mathrm {rad}} } } { \langle
        \psi_{s,\,1/2} | H_{\mathrm {PNC}} |\psi_{p,\,1/2}\rangle~~~ }~.
   \end{equation} 
   The energy difference between the considered opposite parity
   orbitals $\sim 1$ eV is much lower than a typical excitation energy
   $\sim m=0.5$ MeV that governs the radiative corrections.  We can
   therefore neglect this difference assuming that $E_{s,1/2}\simeq
   E_{p,1/2}$.  This assumption makes the correction $\langle
   \psi_{s,\,1/2} | H_{\mathrm PNC}|\psi_{p,\,1/2}\rangle^{\mathrm rad}$
   gauge invariant.  To make this argument more transparent, one can
   use the Coulomb approximation for the atomic field and consider the
   degenerate Coulomb $ns_{1/2}$ and $np_{1/2}$ levels.

        The problem of gauge invariance can be viewed from another
        perspective. One can consider the full PNC
        amplitude. Alongside the PNC matrix element it includes
        also the conventional E1 amplitude and the corresponding
        energy denominator. The PNC amplitude describes the events on
        the mass shell, the photon frequency in the E1 transition
        equals the excitation energy of the atom. Therefore the
        amplitude is definitely gauge invariant.  Considering the
        radiative corrections in the lowest order, one can separate
        them into corrections to the PNC matrix element, the Lamb-shift
        corrections to the energies, and the corrections to the E1
        transition. Their sum is definitely gauge invariant. The point
        is that the latter two corrections are known to compensate
        each other almost completely \cite{dzuba_01}. The only
        remaining corrections are the ones which are associated with
        the PNC matrix element. Thus these latter corrections are
        gauge invariant. To make this discussion complete, let us
        mention also that in the perturbation theory approach outlined
        in the next section \ref{Zalpha} the gauge invariance of the
        PNC matrix element is self-evident. There is no doubt,
        therefore, that one can assume that the radiative corrections to the
        PNC matrix element are gauge invariant.

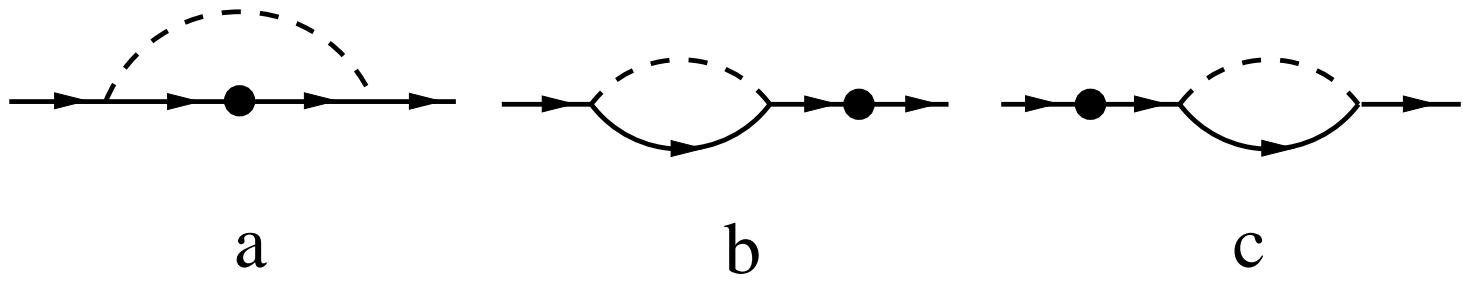
  \begin{figure}[h]
  \caption{  \label{two} 
    The Feynman diagrams for QED vertex (a) and self-energy (b),(c)
    corrections, called e-line corrections in the text. They are used
    in the text to describe either the PNC amplitude or the energy
    shifts induced by the finite nuclear size.  The thick dot
    represents in these two cases either the vertex induced by the PNC
    Hamiltonian, or the finite nuclear size (FNS) potential that takes
    into account the spread of the nuclear charge inside the nucleus.
    The thick line describes the electron propagation in the atomic
    field.  
}\end{figure}

        The quantities $\delta_{ {\mathrm {FNS}},\,s}$ and
        $\delta_{{\mathrm {FNS}},\,p}$ in Eq.(\ref{d}) are the
        relative radiative corrections to the FNS energy shifts
        $E_{\mathrm {FNS},\,s}, ~E_{\mathrm {FNS},\,p},$ for the
        chosen $s_{1/2}$ and $p_{1/2}$ electron states

        \begin{eqnarray}\label{dd} \delta_{ {\mathrm {FNS}},\,l} =
        \frac{ E_{ {\mathrm {FNS}},\,l}^{\mathrm {rad} } }
        { E_{\mathrm {FNS},l } } ~, \quad l = 0,1 ~.  \end{eqnarray}
        The FNS energy shifts can be presented as matrix elements of
        the potential $ V_{\mathrm {FNS}}(r)$ that describes the deviation
        of the nuclear potential from the pure Coulomb one inside the
        nucleus due to the spread of the nuclear charge

        \begin{equation}
          \label{EFNS}
        E_{{\mathrm {FNS}},\,l } = 
        \langle \psi_{ l,\,1/2 } | V_{\mathrm {FNS}}
        |\psi_{l,\,1/2}\rangle~,\quad l=0,1. 
        \end{equation}
        Equality (\ref{ddd}) may be established for the sum of all QED
        radiative corrections ($\sim Z \alpha^2 f(Z \alpha)$), or
        specified for any gauge invariant class of them. We will discuss
        first the self-energy and vertex corrections, restricting our 
        consideration to the lowest order of  perturbation theory
        presented by the Feynman diagrams in Fig.\ref{two}, calling them
        the e-line corrections. Later we will switch our attention to
        the vacuum polarization and demonstrate that it satisfies
        (\ref{ddd}) as well.

        The intermediate electron states in diagrams of Fig. \ref{two}
        are described by the propagator $ G = (\gamma_\mu p^\mu +
        \gamma_0 U - m)^{-1}$, with $p^\mu = (\epsilon, -i {\mbox
          {\boldmath $\nabla$} } )$, where $\epsilon=m+E\simeq m$ is the
        relativistic electron energy, $U=U(r)$ is the atomic potential
        that includes the potential created by the nucleus with FNS and,
        generally speaking, the screening potential of atomic electrons.
        The latter one is not important at small distances, but it still
        can be accounted for by the formalism. For the relativistic
        propagator we use the same symbol $G$ that was applied
        previously for the nonrelativistic Green function in
        Eq.(\ref{delta}), which should not produce confusion.  At small
        distances the exchange potential is negligible, therefore the
        potential is local. The external legs of the diagrams describe
        the wave functions $ \psi_{s,1/2}({\bf r})$ and
        $\psi_{p,1/2}({\bf r})$ for the considered $s_{1/2}$ and
        $p_{1/2}$ levels.  Let us note first that in the region of short
        distances the following relations hold

  \begin{eqnarray}\label{prop}
   \gamma_5  G & = & - G \gamma_5 ~, 
  \\ \label{wf}
  \psi_{s,1/2}({\bf r}) & = & ~~\,c \gamma_5 \psi_{p,1/2}({\bf r})~.
      \end{eqnarray}
      They follow from the chiral invariance that manifests itself at
      small distances. For $r \ll
      1/m$ the mass term in the Dirac equation becomes smaller than the
      kinetic term ($\propto 1/r$). For even smaller distances $ r\le
      Z\alpha/m$ the potential ($Z\alpha/r$) also exceeds the mass
      term. Therefore for small distances we can neglect the mass
      $m\Rightarrow  0$, using the resulting chiral invariance for our
      advantages.
      Since Eqs.(\ref{prop}),(\ref{wf}) will be important for us, we
      examine them in some detail.  In order to prove Eq.(\ref{prop})
      one neglects the mass term in the electron propagator assuming
      that

  \begin{equation} \label{m=0} G^{-1} = \gamma_\mu p^\mu + \gamma_0
    U~.  \end{equation} Then, the identity $\gamma_5 \gamma_\mu = -
    \gamma_\mu \gamma_5$ leads to $G^{-1} \gamma_\mu = - \gamma_\mu
    G^{-1}$ that results in (\ref{prop}).  In order to verify
    Eq. (\ref{wf}) let us write down the conventional Dirac equations
    for the large $f$ and small $g$ components of the electron wave
    function (\ref{spinor})

  \begin{eqnarray}
  \label{dirac1}
  f' +\frac{1+\kappa}{r} f - (\epsilon + m - U) g &=& 0~, 
  \\ \label{dirac2}
  g' +\frac{1-\kappa}{r} g + (\epsilon - m - U) f &=& 0~.
    \end{eqnarray}
    Here we omitted the indexes $njl$, distinguishing the states by the
    parameter $\kappa$ that takes the values $-1$ and $1$ for the
    $s_{1/2}$ and $p_{1/2}$ states respectively.  For small separations
    one can neglect in
    Eqs.(\ref{dirac1}),(\ref{dirac2}) the mass term $m$. After that we
    observe that the equations become invariant under a substitution
    \begin{eqnarray}
  \nonumber
  \kappa &\rightarrow& -\kappa~, 
  \\ \    \label{1-1}
  f &\rightarrow& ~~\,g~,
  \\ \nonumber
  g &\rightarrow& -f~.
     \end{eqnarray}
     Remember now that according to Eq.(\ref{spinor}) the 
       $s_{1/2}$ and $p_{1/2}$ wave functions have the form

     \begin{eqnarray} \label{sp}
       \psi_{s,1/2} =
       \left(
         \begin{array}{cc} f_{s,1/2}\,u \\
         i \, g_{s,1/2}(- \mbox{\boldmath{$\sigma$}} \cdot {\bf n} ) \,u
         \end{array} \right), \\ \nonumber
       \psi_{p,1/2} =
       \left(
         \begin{array}{cc} f_{p,1/2} 
          (- \mbox{\boldmath{$\sigma$}} \cdot {\bf n})\,u \\
            i \,g_{p,1/2}\,u
         \end{array} \right)~.
      \end{eqnarray}
      Here the two component spinor $u = u_\alpha,~\alpha=1,2$ depends
      on the projection of the total momentum $\mu$, $u_\alpha =
      \delta_{\alpha,1}$ for $\mu=1/2$, and $u_\alpha =
      \delta_{\alpha,2}$ for $\mu=-1/2$.  It follows from Eq.(\ref{sp})
      that

  \begin{equation}
    \label{g5p}
    \gamma_5 \,\psi_{p,1/2} = i
         \left( \begin{array}{cc} g_{p,1/2} \,u \\
            -i f_{p,1/2} (-\mbox{\boldmath{$\sigma$}} \cdot {\bf n}) \,u
         \end{array} \right)~,
      \end{equation}
  where we took into account that in the standard
  representation considered for the Dirac matrixes

  \begin{equation}
    \label{g5def}
    \gamma_5 = \left( \begin{array}{cc} 0 & 1\\ 1 &0 \end{array}
  \right)~.
      \end{equation}
      From Eq.(\ref{g5p}),(\ref{sp}) one observes that 
      Eq.(\ref{wf}) can be presented as 

      \begin{eqnarray} \label{fg} f_{s,1/2}&=& ~~\,c g_{p,1/2}~, \\
        \label{gf} g_{s,1/2} &=& -c f_{p,1/2}~.
      \end{eqnarray} 
      We see that these latter relations coincide with 
      Eq.(\ref{1-1}). The first substitution $\kappa \rightarrow
      -\kappa$ in (\ref{1-1}) exchanges equations for the $s_{1/2}$ and
      $p_{1/2}$ states, while the two others are identical to
      Eqs.(\ref{fg}) and (\ref{gf}) respectively. Thus the verified
      above Eqs.(\ref{1-1}) prove validity of Eq.(\ref{wf}).

      At this point one may recall a subtlety, a symmetry of the Dirac
      equation revealed by Eqs.(\ref{1-1}) is, generally speaking, not
      sufficient. One needs to verify also that the
      boundary conditions satisfy same symmetry as well. However, one
      can argue that the boundary conditions arise as a physical
      complement to the Dirac equation and, therefore, the symmetry
      of the equation should automatically result in the symmetry of the
      boundary conditions.  To verify this argument one can recall that
      for the finite nuclear charge distribution the boundary conditions
      at the origin read simply

      \begin{equation}
        \label{bc}
        \left\{ \begin{array}{ll}
  f_{s,1/2}(0) = const~, & \\
  g_{s,1/2}(0) = 0~, 
  \end{array}  \right.
  \quad 
  \left\{ \begin{array}{ll}
  f_{p,1/2}(0) = 0~, & \\
  g_{p,1/2}(0) = const~, 
  \end{array} \right.
      \end{equation}
      where $ const \ne 0$. Eqs.(\ref{bc}) explicitly demonstrate that
      indeed, the boundary conditions are invariant under the
      substitution Eq.(\ref{1-1}). This completes our verification of
      Eq.(\ref{wf}).  The constant $c$ in (\ref{wf}), which depends on
      the normalization conditions, will be irrelevant for us because we
      are interested in relative quantities that appear in
      Eq.(\ref{ddd}) where this constant will be canceled out.
      
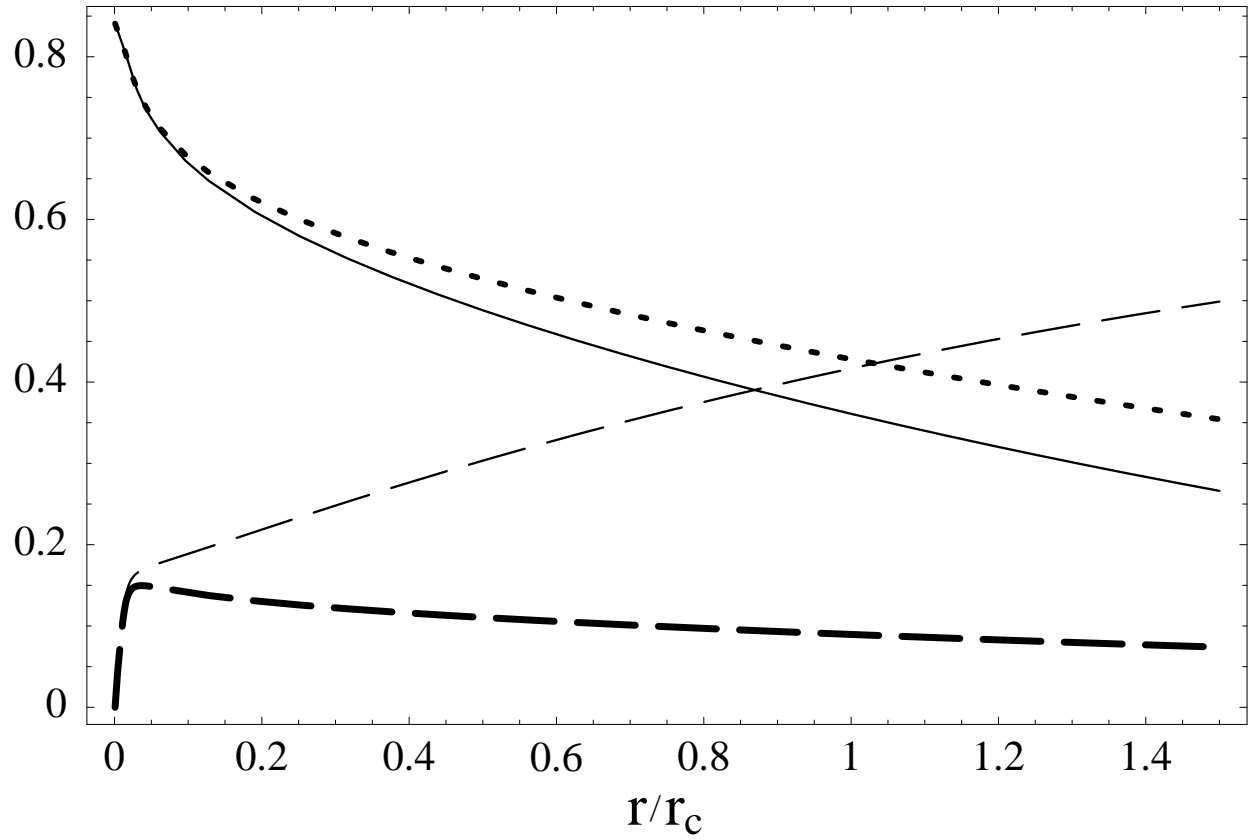
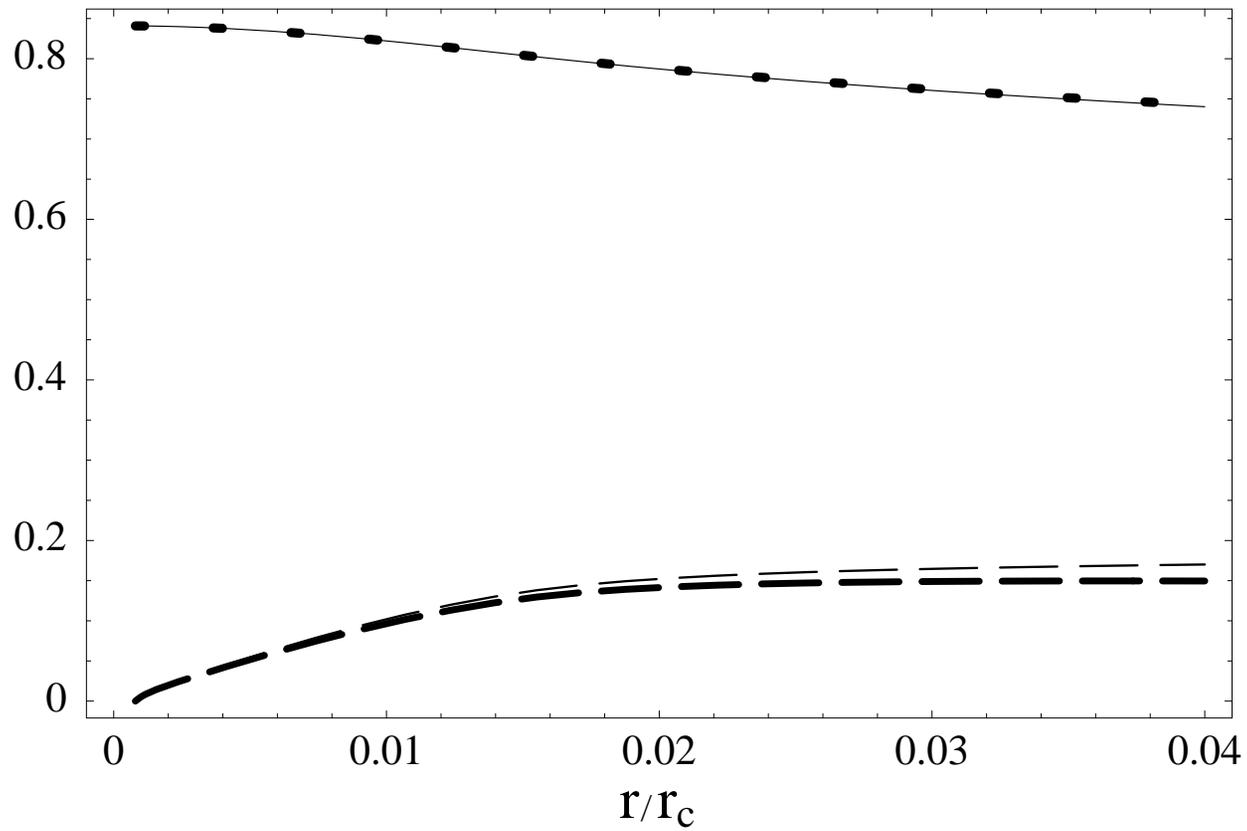
  \begin{figure}[h] 
  \caption{\label{three}
    The $s_{1/2}$ and $p_{1/2}$ wave functions for the valence
    electron in $^{133}$Cs (arbitrary units) versus the distance
    $r/r_\mathrm{C}$, where $r_\mathrm{C}=\hbar/mc$ is the Compton
    radius.  The functions are normalized in such a way that the
    $const$ in both sets of boundary conditions in Eq.(\ref{bc}) takes
    the same value, $const = 1$. Solid and dotted lines - $f_{s,1/2}$
    and $g_{p,1/2}$ wave functions that are large for this
    normalization, thick dashed and thin dashed lines $-g_{s,1/2}$ and
    $f_{p,1/2}$ wave functions that are much smaller, (a) the region
    of large distances $r\sim r_\mathrm{C}$ shows good coincidence of
    $f_{s,1/2}$ with $g_{p,1/2}$, (b) the region in the vicinity of
    the nucleus, the nuclear radius is $r_\mathrm{N} \simeq
    0.016~r_\mathrm{C} \simeq 6.1$ fm, where $-g_{s,1/2}$ coincides
    with $f_{p,1/2}$.  The shown results illustrate Eq.(\ref{wf})
    based on chiral invariance.  }\end{figure}

      It is instructive to look at numerical verifications of
      Eq.(\ref{wf}) shown in Fig.\ref{three}.  Supporting the proof
      given above, they provide also details that are cumbersome for
      analytic treatment.  Fig.\ref{three} (a) shows that
      Eq.(\ref{fg}), which according to the discussion above holds at
      small distances $r\ll 1 /m$, remains approximately valid at
      sufficiently large distances, up to (and even beyond) the
      Compton radius $r \le 1/m$.  In contrast, Eq.(\ref{gf}) is
      applicable only at small $r \ll 1/m$ as shown in Fig.
      \ref{three} (b), being violated at $ r \sim 1/m $, as one can
      see from Fig. \ref{three}(a). We can remember, however, that the
      dominant contribution to the PNC matrix element arises from the
      $f_{s,1/2}$ and $ g_{p,1/2}$ functions, which are both large for
      $r\le 1/m$, much larger than the other two functions
      $g_{s,1/2},f_{p,1/2}$.  Since these large functions satisfy
      Eq.(\ref{fg}) even at sufficiently large distances $r \le 1/m$,
      one can also rely in this region on Eq.(\ref{wf}).  We will
      exploit this opportunity at some later stage below, but our
      prime interest is related to the distances as small as the
      nuclear radius $r \le r_\mathrm{N}\ll Z\alpha/m$, where we
      verified beyond doubt that Eqs.(\ref{prop}),(\ref{wf}) hold, see
      Fig.\ref{three} (b).

      Consider the diagrams (b),(c) in Fig. \ref{two} that describe the
      self-energy correction for the PNC amplitude.  The thick lines in
      this figure represent the electron propagation in the atomic
      field, the dot vertex is the PNC interaction described by
      Eq.(\ref{05}), and the dashed line shows the photon.  Let us
      specify that the left and right legs of these diagrams describe
      the $s_{1/2}$ and $p_{1/2}$ states respectively. The PNC
      interaction (\ref{05}) is located at small distances $r_\mathrm{
        N} \ll 1/m $.  Let us use this fact and, relying on
      Eqs.(\ref{prop})(\ref{wf}), transform the PNC matrix element. To
      simplify the notation it is convenient to present the Hamiltonian of
      the weak interaction Eq.(\ref{05}) in the form

     \begin{equation}\label{005} H_{\mathrm {PNC}} = V_\mathrm{PNC}
     (r)\,\gamma_5 ~, 
     \end{equation} 
     where $V_\mathrm{PNC} (r) = G_{\mathrm F} \, Q_{\mathrm W}\,
     \rho(r)/(2 \sqrt{2})$ is a scalar factor in the PNC interaction.
     Consider first the diagram (b). The $\psi_{p,1/2}({\bf r}) $
     function in this diagram enters the PNC interaction that makes its
     argument ${\bf r}$ localized inside the nucleus, $r\le
     r_\mathrm{N}$. This allows us to use Eq.(\ref{wf}) deriving

  \begin{eqnarray}\label{se(b)}
  \langle \psi_{p,\,1/2} | \, \gamma_5  \,V_\mathrm{PNC}\,  G \,\Sigma\,
    |\psi_{s,\,1/2}   \rangle 
\\
\nonumber
\equiv 
  \langle \psi_{s,\, 1/2} | \, V_\mathrm{PNC} \,G \, \,\Sigma  |
\psi_{s,\,1/2}
  \rangle ~. 
     \end{eqnarray}
     Here and below we use the symbol of identity ($\equiv$) that simply
     means dropping the irrelevant constant $c$ introduced in
     Eq.(\ref{prop}) (that is canceled out when the relative contribution
     of the correction is considered anyway).  Eq.(\ref{se(b)}) uses 
     conventional notation, $\Sigma$ and $G$ are the self-energy
     operator and the electron propagator mentioned above.  The
     right-hand side of this relation shows that we can look at the diagram
     (b) as the one that describes the diagonal $s_{1/2}-s_{1/2}$
     transition induced by an effective scalar potential
     $V_\mathrm{PNC}$. Consider now the same diagram (b), but for
     another physical quantity, for the FNS energy shift for the
     $s_{1/2}$ level. Its contribution to the energy reads

  \begin{equation}\label{se(b)fns}
  \langle \psi_{s,\,1/2} | \, V_\mathrm{FNS}\, G \, \Sigma \, |
\psi_{s,\,1/2}
    \rangle ~.  
    \end{equation} 
    Eqs.(\ref{se(b)}),(\ref{se(b)fns}) give the contribution of one and
    the same diagram (b) to different quantities, namely the PNC
    amplitude and FNS energy shifts. Comparing them one
    observes their similarity.  In both cases the diagram describes the
    $s_{1/2}-s_{1/2}$ transitions that are induced by the short-range
    scalar potentials.  The only distinction comes from the fact that
    the potentials $V_\mathrm{PNC}(r)$ and $V_\mathrm {FNS}(r)$ possess
    different profiles.  

    This latter difference, however, proves to be irrelevant.  In order to
    see this we need to discuss the integral with a short-range
    potential that arises in the diagram.  Its integrand includes the
    potential (i. e.  $V_\mathrm{PNC}(r)$ for the PNC amplitude and
    $V_\mathrm{FNS}(r)$ for the FNS energy shift), the electron wave
    function $\psi_{s,1/2}({\bf r})$ and the electron propagator $G({\bf r},
    {\bf r}')$.  The wave function and the propagator inside the nucleus
    are almost constants, see discussion of this fact below.  Taking
    them out of the integration we observe that the considered
    sophisticated integral is simplified, being reduced to the
    integral over the potential itself $\int V_\mathrm{PNC}(r) d{\bf r}$.
    It is easy to see that the same integral arises in the non-perturbed
    matrix element.  Therefore in the relative contribution of the
    correction this integral is canceled out, and the result does not
    depend on the potential at all.  This conclusion is valid for any
    short-range potential.  Thus the relative contribution of the
    diagram (b) proves to be one and the same for the PNC and the FNS
    problems. This means that the relative contribution of the
    diagram (b) to the PNC matrix element equals half of its relative
    contribution to the energy shift of the $s_{1/2}$ state due to FNS.
    The mentioned factor $1/2$ takes into account the fact that the 
    energy shift is induced by the two diagrams (b) and (c) in Fig.
    \ref{two} that are identical for the FNS problem.
    
    Before proceeding further, let us discuss the behavior of the
    electron wave functions inside the nucleus which was exploited in
    the above derivation.  Note that inside the nucleus the large
    functions $f_{s,1/2}(r)$ and $g_{p,1/2}(r)$ exhibit smooth, almost
    constant behavior, as demonstrates Fig. \ref{three}.  The reason
    for this effect stems from the fact that inside the nucleus the
    electric field rapidly diminishes with the radius. The functions
    $g_{s,1/2}(r)$ and $f_{p,1/2}(r)$, show some variation, but it
    remains very mild. Assessing consequences of this smooth variation
    one can recall that, firstly, the integration volume $r^2dr$
    enhances the contribution of the nuclear surface region where
    these functions are very smooth indeed, as shows Fig.  \ref{three}
    (b).  Secondly, their role is not prominent anyway, simply because
    they are smaller than the functions $f_{s,1/2}(r)\,g_{p,1/2}(r)$.
    Therefore they can be well approximated by constants.  Thus both
    $s_{1/2}$ and $p_{1/2}$ states are well described by constant-type
    functions inside the nucleus.  This fact was used above for the
    $s_{1/2}$ state, the argument is repeated below for the $p_{1/2}$
    state.  We also presumed above that the propagator $G({\bf r},
    {\bf r}')$ does not depend on the variable ${\bf r}$, when $r\le
    r_\mathrm{N}$.  This statement follows from the simple fact that
    the coordinate ${\bf r}'$ gives the location of the point where
    the radiation of the photon takes place, therefore $r'\sim 1/m \gg
    r$.  The smallness of $r$ compared with $r'$ justifies its neglect
    in the propagator.

    Examining the diagram Fig.\ref{two} (c) we use the same arguments
    as we exploited above for the (b) diagram proving that its
    relative contribution to the PNC amplitude equals half of its
    relative contribution to the energy shift due to FNS of the
    $p_{1/2}$ state.  Combining our results for the diagrams (b) and
    (c) we conclude that the self-energy corrections (superscript SE)
    described by these diagrams comply with Eq.(\ref{ddd})

     \begin{equation}
       \label{ds}
             \delta_{\mathrm {PNC},\,sp} ^\mathrm {(SE)} = \frac{1}{2}\,  
        \Big( \, \delta_{ \mathrm {FNS},\,s }^\mathrm {(SE)} +
        \delta_{\mathrm {FNS},\,p }^\mathrm {(SE)}\, \Big)~.
     \end{equation}
     The given derivation does not use any specific gauge condition,
     therefore it is gauge invariant.

     Consider now the vertex correction (a) in Fig. \ref{two}.  An
     analytical expression for it can be written in the conventional
     form
     \begin{eqnarray}
       \label{vertex}
- i\alpha  \int \!\! \frac{d^4q}{(2\pi)^4}   
d{\bf r} d{\bf r}'  
  d{\bf r}_0
   D^{\mu\nu}(q) 
\\
\nonumber
\times
  e^ {i \,{\bf q}\cdot ({\bf r} -{\bf r}')} 
   \Phi_{\mu\nu}({\bf r}, {\bf r}', {\bf r}_0)~,
  \end{eqnarray}
  where 
  \begin{eqnarray} 
\nonumber  
\Phi_{\mu\nu}({\bf r},  {\bf r}', {\bf r}_0) &=&
  \bar    \psi_{p,1/2}({\bf r})  \gamma_\mu 
  G({\bf r},{\bf r}_0; \epsilon + \omega)
  \\   \nonumber
  &\times&
  V_\mathrm{PNC}(r_0)  
\gamma_0 \,\gamma_5 \,
  G({\bf r}_0 , {\bf r}'; \epsilon + \omega) 
\\ 
\label{Phi}
&\times&
\gamma_\nu 
  \psi_{s,1/2}({\bf r}')~.
     \end{eqnarray}
     Here $D^{\mu\nu}(q)$ is the photon propagator, $q^\mu = (\omega,
     {\bf q})$ is the four-vector of the photon momentum.  The range of
     distances $r_{\mathrm rad}$ where the radiation processes take
     place is of the order of the Compton radius, $r_{\mathrm rad}\sim
     m^{-1}$. This ensures that $ r,r'\gg r_0$ because the PNC
     interaction is localized at $ r_0 \le r_\mathrm {N}$.  We can
     simplify therefore Eq.(\ref{vertex}) factorizing the interaction
     over ${\bf r}_0$

     \begin{eqnarray}
\nonumber
  &&  \int d{\bf r}_0 \,\Phi_{\mu\nu}({\bf r},  {\bf r}', {\bf r}_0)= 
     \int V_\mathrm{PNC} (r_0)\, d{\bf r}_0  \\        \label{r0}
  &&  \times \, \bar    \psi_{p,1/2}({\bf r}) \, \gamma_\mu \,
  G({\bf r},\,{\bf 0}; \,\epsilon + \omega)  \\ \nonumber
&& \times \,\gamma_0 \,\gamma_5 \,
  G({\bf 0} , \,{\bf r}';\, \epsilon + \omega) \,\gamma_\nu \,
  \psi_{s,1/2}({\bf r}')~,
     \end{eqnarray}
     where we put $r_0=0$ in the arguments of the propagators.  We
     observe that the potential arises only in a factor $\int
     V_\mathrm{PNC} (r_0)\, d{\bf r}_0$ that is separated from all other
     elements of the diagram.  The same factor appears in the main term of
     the PNC amplitude.  Therefore the relative contribution of the
     vertex correction does not depend on the potential at all.  Remember
     that the same property is exhibited by the self-energy
     correction discussed above.

     The variables $r,r'$ describe the radiation processes and therefore
     are expected to be of the order of the Compton radius $r,r'\sim
     1/m$. Let us consider the implications that arise if one
     presumes that the region of small separations

     \begin{equation}
       \label{rr}
   r,r' < \frac{1}{m}
       \end{equation}
       gives the dominant contribution to the integral in
       Eq.(\ref{vertex}).  We can engage in this region
       Eqs.(\ref{prop}),(\ref{wf}).  The first of them allows one to
       exchange the $\gamma_5$ matrix in Eq.(\ref{r0}) with the
       propagator $G$. One can also transpose it with the vertexes
       $\gamma_5 \gamma_\mu = - \gamma_\mu \gamma_5 $.  Using several
       such transpositions we can bring the $\gamma_5 $ matrix from
       its central position in expression (\ref{r0}) to the side,
       where it hits the external wave function.  After that we use
       Eq.(\ref{wf}) for this function. Repeating the procedure twice,
       one time shifting $\gamma_5$ to the left, and another one to
       the right we derive

  \begin{eqnarray}
  \nonumber
  && \!\! \bar \psi_{p}({\bf r}) \, \gamma_\mu \,
  G({\bf r},\,{\bf 0}) \, \gamma_0 \,\gamma_5 \,
  G({\bf 0} , \,{\bf r}') \,\gamma_\nu \,
  \psi_{s}({\bf r}') \\ \label{l}
  &&\!\!  \equiv \bar \psi_{s}({\bf r}) \, \gamma_\mu \,
  G({\bf r},\,{\bf 0} ) \, \gamma_0 \,
  G({\bf 0} , \,{\bf r}') \,\gamma_\nu \,
  \psi_{s}({\bf r}') \\ \label{r} 
  &&\!\!   \equiv \bar \psi_{p}({\bf r}) \, \gamma_\mu \,
  G({\bf r},\,{\bf 0} ) \, \gamma_0 \,
  G({\bf 0} , \,{\bf r}') \,\gamma_\nu \,
  \psi_{p}({\bf r}') ~.
      \end{eqnarray}
      To simplify notation we dropped here the total momentum $1/2$
      from the indexes in the wave functions (i. e. $\psi_p \equiv
      \psi_{p,1/2}$ etc), and omitted the energy variable in the
      propagators (this variable equals $\epsilon+\omega$ for all
      propagators).  The point of these transformation is that the
      $\gamma_5$ matrix disappears from the right-hand sides of
      Eqs.(\ref{l}),(\ref{r}).  Substituting these two expressions
      back in Eqs.(\ref{r0}), (\ref{vertex}) we find that the relative
      contribution of the vertex diagram (a) to the PNC matrix element
      can be expressed either as a matrix element of the
      $s_{1/2}-s_{1/2}$ transition, or the matrix element of the
      $p_{1/2}-p_{1/2} $ transition.  We recognize in these two matrix
      elements the relative contributions to the energy shifts due to
      FNS for the $s_{1/2}$ and $p_{1/2}$ states respectively. This
      latter conclusion takes advantage of the fact discussed above
      that the profile of the short-range potential does not influence
      the relative correction, one can conveniently choose the
      potential to be $V_\mathrm{PNC}$ or $V_\mathrm {FNS}$.  This
      discussion demonstrates that the vertex (superscript V)
      corrections to the PNC amplitude and to the energy shifts due to
      FNS prove to be approximately equal 

     \begin{equation}
       \label{V}
             \delta_{\mathrm {PNC},\,sp} ^\mathrm {(V)} \cong
        \delta_{ \mathrm {FNS},\,s }^\mathrm {(V)} \cong
        \delta_{\mathrm {FNS},\,p }^\mathrm {(V)}~.
     \end{equation}
     Combining Eqs.(\ref{ds}),(\ref{V}) we find that the total e-line
     corrections, i. e. the self-energy plus the vertex correction,
     satisfy Eq.(\ref{ddd}). 
     
     The derivation presented above exploits Eq.(\ref{rr}) presuming
     that the small distances are dominant in the vertex correction.
     There are reasons indicating that this assumption does not put
     restrictions on the derived result (\ref{ddd}). As a first
     attempt one can try to exploit the behavior of the wave functions
     $f_{s,1/2},g_{p,1/2}$.  Fig.  \ref{three} shows that they
     approximately satisfy Eq.(\ref{wf}) even at sufficiently large
     separations, of the order of the Compton radius $ r \le 1/m$.
     This shows that Eq.(\ref{l}) remains approximately applicable
     even if the integration region includes sufficiently large
     separations $ r \sim 1/m$.  This, in turn, demonstrates that the
     first equality in Eq.(\ref{V}), i. e. $ \delta_{\mathrm
       {PNC},\,sp} ^\mathrm {(V)} = \delta_{ \mathrm {FNS},\,s
     }^\mathrm {(V)}$ also remains approximately valid. This is a
     positive result.  However, this fact alone is not sufficient.
     The second equality in Eq.(\ref{V}) proves to be much more
     sensitive to the essential integration region because the wave
     functions $f_{p,1/2},~g_{s,1/2}$ fail to satisfy Eq.(\ref{wf}) at
     $r \sim 1/m$. We need therefore to be more careful.

     Let us recall at this stage that the range of distances $r_{\mathrm
       rad}$ where the radiation processes take place depends on the
     chosen gauge for the electromagnetic field.  It is well known that
     the usual length-gauge favors larger distances, while the velocity
     and, especially, acceleration forms, make the smaller distances 
     contribute more to the radiation process. These known examples
     show that the region that contributes to the radiation
     process is {\em not} gauge invariant.  This is important.  If the
     radiation region is not gauge invariant, then we are able to choose
     a gauge in which this region is located close to the nucleus,
     inside the zone defined by Eq.(\ref{rr}). In this gauge the above
     derivation is valid and Eq.(\ref{V}) for the vertex is correct.
     Note that  this argument does not rely on an explicit form for
     the necessary gauge. It suffices to acknowledge only that the
     radiation region is not gauge invariant. Summarizing, there exists
     a gauge (more accurately, a family of gauges) in which
     Eq.(\ref{V}) holds.

     Recall now that Eq.(\ref{ds}) for the self-energy remains valid
     in any gauge, in particular in the one discussed above. In this
     gauge we can combine together the self-energy and vertex
     corrections (\ref{ds}),(\ref{V}) proving the validity of their sum
     Eq.(\ref{ddd}).  The three quantities in this latter relation are
     all well-defined, gauge invariant physical observables. The fact
     that Eq.(\ref{ddd}) is derived in one particular gauge means that
     it is valid in any gauge, as was proposed in \cite{kf_prl_02}.

     Let us summarize the above discussion.  We argued that the
     calculations can be organized in such a way that all important
     events happen at small distances. In this region the chiral
     invariance holds, resulting in Eqs.(\ref{prop}),(\ref{wf}) and,
     consequently, in Eq.(\ref{ddd}). Thus this latter equality
     expresses the fundamental and simple fact. At small, nuclear
     distances the chiral invariance governs the problem.
     Consequently, Eq.(\ref{ddd}) that follows from it, can be called
     the chiral invariance identity.

     Let us estimate the accuracy of Eq.(\ref{ddd}).  The above
  derivation used the gauge in which the radiation processes for the
  diagram (a) take place mostly at small distances $r_\mathrm {rad} <
  m^{-1}$.  They should also take place outside the nucleus $r_\mathrm
  {N}<r_{\mathrm rad}$, as is necessary to justify our presumption
  that the shape of the potential inside the nucleus is irrelevant.
  From the last two inequalities we find that the derivation relies on
  a parameter $\xi = mr_{\mathrm N} \sim 0.01$.  This determines the
  magnitude of the error (few per cent) of Eq.(\ref{ddd}).

  Similarly one considers the contribution of the QED vacuum
  polarization.  Eqs.(\ref{final}),(\ref{a}) and (\ref{k}) present
  explicit variations for $s_{1/2}$ and $p_{1/2}$ wave functions at the
  origin induced by the vacuum polarization. Using these wave functions
  to calculate corrections to the PNC matrix element and FNS energy
  shifts, one immediately finds that the chiral invariance identity
  (\ref{ddd}) holds for the vacuum polarization as well.  This fact is
  in line with Eq.(\ref{ds}) for the self-energy corrections.  The proof
  of (\ref{ddd}) for the vacuum polarization stops here because one need
  not care about the complicated vertex corrections.

  Note that we do not consider above the radiative corrections of the
  order $\sim \alpha/\pi$ which appear in the plane wave approximation.
  These contributions have been included into the radiative corrections
  to the weak charge $Q_W$ (and the renormalization of the charge and
  electron mass in the case of FNS energy shifts).  Correspondingly, we
  subtract the contribution of the plane waves from Eq.(\ref{ddd}),
  considering only the part of the corrections that depends on the
  atomic potential $\sim Z\alpha^2 f(Z\alpha)$.  For heavy atoms this
  subtlety is insignificant numerically because the considered
  $Z$-dependent part of the correction is bigger than the omitted
  $Z$-independent one, as we will see below.

  The chiral invariance identity (\ref{ddd}) dives the e-line
  corrections to the PNC matrix element, which are difficult to
  calculate, in terms of the corrections to the FNS energy shifts that
  have been well-studied both numerically, by Johnson and Soff
  \cite{johnson_soff_85}, Blundell \cite{blundell_92}, Cheng {\em et
  al} \cite{cheng_93} and Lindgren {\em et al} \cite{lindgren_93}, and
  analytically, by Pachucki \cite{pachucki_93} and Eides and Grotch
  and Eides {\it et al} \cite{eides_97}.  Ref.  \cite{cheng_93}
  presents the e-line radiative corrections to the FNS energy shifts
  for $1s_{1/2}$, $2s_{1/2}$ and $2p_{1/2}$ levels in hydrogenlike
  ions with atomic charges $Z=60,70,80,90$.  Eq.  (\ref{ddd}) contains
  relative corrections, therefore we needed to calculate the FNS
  energy shifts $E_{\mathrm {FNS}}$. This was done in \cite{kf_prl_02}
  by solving the Dirac equation with the conventional Fermi-type
  nuclear distribution $\rho(r) = \rho_0 /\{ 1 + \exp [(r-a)/c] \} $.
  Parameters $a,c$ were taken the same as in \cite{cheng_93}, namely
  $a = 0.523$ fm and $c$ chosen to satisfy $R_{\mathrm {rms}} = 0.836
  A^{1/3} + 0.570$ fm.
    
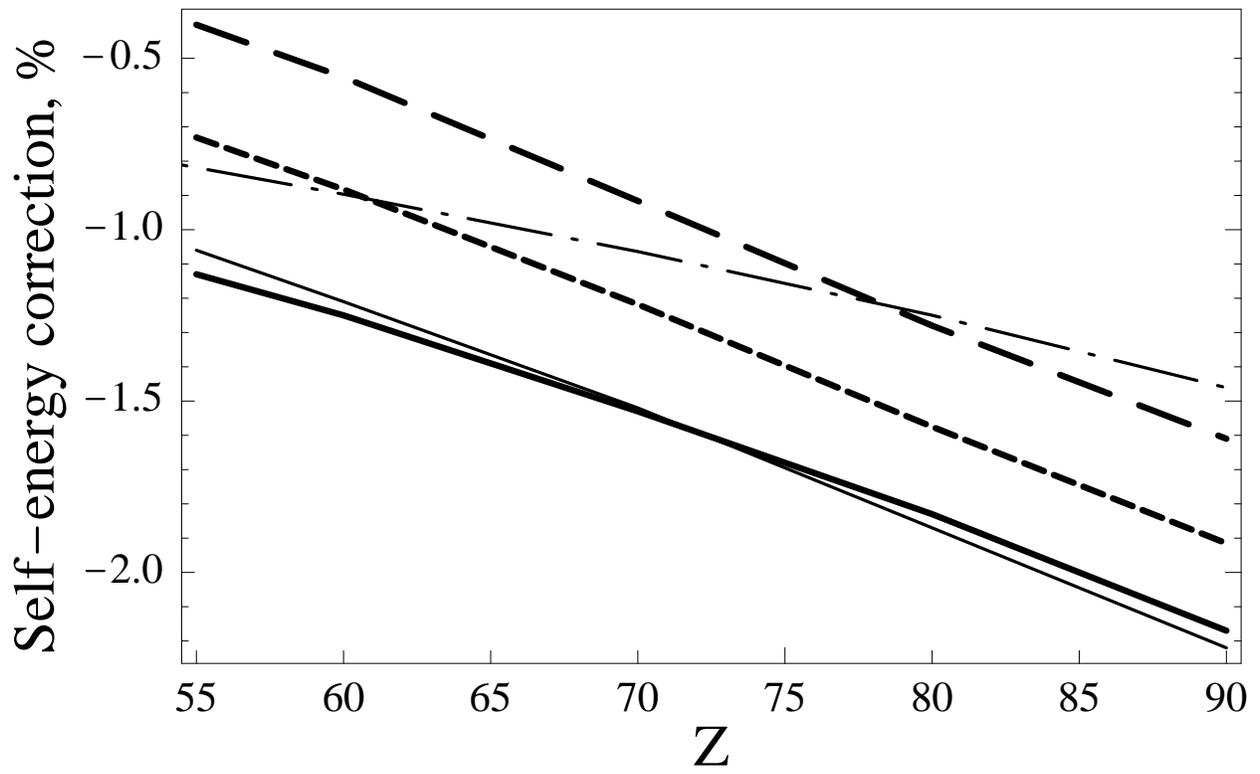
  \begin{figure}[h]
  \caption{\label{four} 
    The relative radiative corrections ($\%$) induced by the e-line
    corrections (self-energy plus vertex) of Fig.\ref{two}.  Thick,
    thin, and long-dashed lines - corrections for FNS energy shifts
    for $1s_{1/2}$, $2s_{1/2}$, and $2p_{1/2}$ levels respectively
    extracted from \cite{blundell_sapirstein_johnson_92}; short-dashed
    line - prediction of Eq.(\ref{ddd}) for the PNC matrix element
    calculated in \cite{kf_prl_02}, dashed-dotted line - correction to
    the hyperfine interaction \cite{kf_prl_02}.  
}\end{figure}

  Using the results of \cite{cheng_93} and this calculation we
  obtained in \cite{kf_prl_02} the relative radiative corrections
  shown in Fig.  \ref{four}.  In order to include the interesting case
  $Z=55$ and to account for all values of $55 \le Z \le 90$ we used
  interpolating formulae presented in \cite{cheng_93}, as well as data
  of \cite{cheng_02}. The relative corrections for the $1s$ and $2s$
  levels are approximately the same size. This indicates that the
  radiative processes responsible for the correction take place at
  separations much smaller than the K-shell radius, $r \ll (Z \alpha
  m)^{-1}$, which is consistent with the assumption $r \le 1/m$ above.
  For these separations we can assume that, firstly, the screening of
  the nuclear Coulomb field in manyelectron atoms does not produce any
  significant effect, and, secondly, the relative corrections do not
  depend on the atomic energy level because for small separations all
  atomic $ns_{1/2}$-wave functions exhibit similar behavior.  These
  arguments remain valid for the $p_{1/2}$ states as well, permitting
  us to presume that the results shown in Fig.  \ref{four} for the
  $2s_{1/2}$-levels and $2p_{1/2}$-levels of hydrogenlike ions remain
  valid for $s_{1/2}$ and $p_{1/2}$ states of the valence electron in
  a manyelectron atom. We obtain the e-line radiative corrections for
  the PNC matrix element using the identity (\ref{ddd}) that expresses
  them via the found corrections to the FNS energy shifts.  The found
  PNC corrections, presented in Fig.  \ref{four} by the dotted line,
  are negative and large.  For the $^{133}$Cs atom the correction is
  $-0.73(20) \%$, for Tl it is $-1.6 \% $. The result for $^{133}$Cs
  is important because it indicated for the first time that the
  radiative corrections reconcile the experimental data of Wood {\em
    et al} \cite{wood_97} with the standard model \cite{kf_prl_02}.
  
  A notable feature of the self-energy corrections presented in Fig.
  \ref{four} is their close similarity. The corrections to the FNS
  energy shifts for both $s_{1/2}$ and $p_{1/2}$ states, as well as
  for the PNC amplitude, have large negative contributions from the
  QED self-energy, which monotonically and very smoothly increase with
  the nuclear charge. This observation brings us back to the point
  mentioned in section \ref{intro}. The large and negative
  contribution of the self-energy for heavy atoms is exactly what one
  {\em should} expect from it. Anything else would be a surprise. In
  order to verify this point one can try to apply the approach outlined
  to some other problem, for example to the hyperfine interaction HFI
  which has been carefully examined previously, see detailed
  discussions in \cite{blundell_97,sunnergren_98} and references
  therein \footnote{ One of us (M.K.) is thankful to A.I.Milstein for
    the second reference.}.

  With this purpose one can try to derive the chiral invariance
  identity for the HFI.  The problem is that the HFI has a long-range
  tail $\sim 1/r^3$ that is prominent in the region where the chiral
  invariance is violated. However, if one believes that convergence of
  the HFI matrix elements is fast enough, then the following equality

  \begin{equation} \label{hfi} \delta_{ {\mathrm {FNS}},\,s} \approx
    \delta_{ {\mathrm {HFI}},\,s}'~, 
    \end{equation} 
    that was proposed in \cite{kf_prl_02} should hold at least
    approximately. Here $\delta_{ {\mathrm {HFI}},\,s}'$ is the
    radiative correction to the HFI for $s_{1/2}$-levels, the primed
    notation indicates that the $Z$-independent Schwinger term
    $\alpha/(2\pi)$ should be excluded.  For heavy atoms this subtlety
    is not important, since the contribution from the Coulomb-induced
    corrections is much stronger than the $Z$-independent Schwinger
    term.  Fig.\ref{four} shows the e-line contribution to $\delta_{
      {\mathrm {HFI}},\,s}'$ that was extracted in \cite{kf_prl_02}
    from data of \cite{cheng_93} using interpolation for all values of
    $Z$ considered there.  It agrees semi-quantitatively with
    Eq.(\ref{hfi}), the deviation is less than 33 \%. We observe again the
    same trend, the self-energy correction is large and negative, in
    accordance with the clear physical reasons mentioned in section
    \ref{intro}.  Another positive conclusion is that chiral
    invariance identities similar to (\ref{hfi}) can be
    applicable even for those potentials that spread out of the
    nuclear core.

    The approach outlined raises two points.  Firstly, it is
    interesting to have an estimate for the self-energy correction in
    simple terms, that would not appeal to sophisticated numerical
    calculations.  Secondly, it is important to derive the self-energy
    corrections for lighter elements.  Direct numerical calculations
    are difficult for them because the error rapidly increases for
    lighter atoms. The main source for the error is the FNS energy
    shift for the $p_{1/2}$ state.  The lightest element for which the
    self-energy corrections have been calculated is Cs, where the
    error for the FNS energy shift in the $p_{1/2}$ wave was $\simeq
    100 \% $ \cite{cheng_02}.  Exactly this error results in the
    uncertainty of the above mentioned result $-0.73(20) \% $. For
    heavier atoms the numerical errors of \cite{cheng_93} rapidly
    decrease.  Therefore the main error in our result for very heavy
    ($Z \ge 80$) atoms is associated with the validity of
    Eq.(\ref{ddd}), that was estimated above as a few percent.

     The estimate for the magnitude of the effect and calculations for
     lighter elements can be conveniently performed with the help of the
     $Z \alpha$-expansion considered in the next section.

  \subsection{ Perturbation theory in powers of  
    $Z \alpha $  }
    \label{Zalpha}
    
    Let us consider the perturbation theory for the e-line
    corrections.  In the initial approximation one describes the
    electron using plane waves and consequently including in the
    perturbation theory three types of processes. Firstly, the PNC
    interaction, secondly, the e-line corrections, and, thirdly, the
    Coulomb interaction.  Taking the nuclear Coulomb field as a
    perturbation one can formulate the perturbation theory in powers
    of the nuclear charge $Z$. The parameter that governs the
    corresponding expansion is $\alpha (Z \alpha)^n$, where $n$ is the
    order of the perturbation theory. This statement follows from a
    conventional, well known result \cite{LLIII}.  In the Coulomb
    problem the perturbation runs in powers of $Z e^2/\hbar v$, where
    $v$ is a typical velocity of the electron. For short distances $
    v\simeq 1$, thus $Z e^2/\hbar v = Z \alpha$. The type of expansion
    considered typically includes some additional large logarithmic
    factors, as specified below.  In the lowest order of perturbation
    theory the simplest, linear in $Z$ correction is $\propto
    Z\alpha^2 $. Thus formulated problem requires calculation of the
    relevant coefficients that can be achieved using the corresponding
    Feynman diagrams.

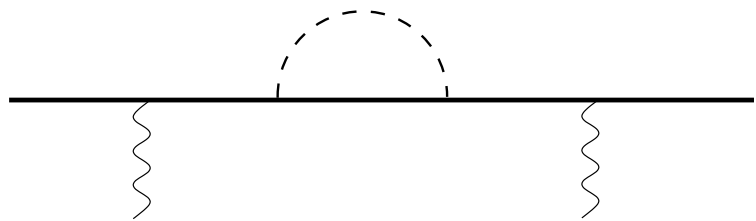
  \begin{figure}[h]
  \caption{ \label{five}
    The lowest order Feynman diagrams for the QED self-energy and
    vertex corrections, called e-line corrections in the text, to the
    PNC matrix element.  The thin line describes propagation of a free
    electron.  For each diagram one of the wavy legs shows the Coulomb
    interaction with the nucleus, another one - the weak PNC
    interaction with the nucleus (which was shown by a large dot in
    Fig. \ref{two} ).  Each diagram represents all possible Feynman
    diagrams with the given topological structure.  
}\end{figure}

    The Feynman diagrams for the e-line corrections in the first order
    of perturbation theory over the Coulomb field are presented in
    Fig. \ref{five}. The direct calculation of these diagrams for the PNC
    amplitude gives the following result for the relative contribution
    of the e-line corrections to the PNC amplitude 

    \begin{equation} \label{197} \delta_{\mathrm{PNC},sp}^
      { (\mathrm{e-line},\,1)} = - 1.97 \,Z\alpha^2~.
   \end{equation}
   It was first obtained in \cite{k_jpb_02} that had an amusing
   history \cite{history}.  The index 1 in the superscript of
   (\ref{197}) reminds one that this is the linear in $Z$ correction.
   For the $^{133}$Cs atom Eq.(\ref{197}) gives $-0.6 \% $, which is
   in line with the result $- 0.73(20) \%$ that follows from the chiral
   invariance identity (\ref{ddd}) as discussed in the previous
   section.  An earlier attempt of Milstein and Sushkov
   \cite{milstein_sushkov_01} gave the coefficient $\approx \!  +0.1$,
   instead of $-1.97$ in Eq.(\ref{197}).  This distinction had
   important implications. Based on their result the authors of
   \cite{milstein_sushkov_01} claimed that there existed the
   contradiction between the experimental data of \cite{wood_97} and
   the standard model, while Eq.(\ref{197}) indicated that the
   experimental data agree with the model.  Fortunately, the
   controversy was soon resolved by Milstein {\it et al} in the
   following Ref.\cite{milstein_sushkov_terekhov_02} that agreed with
   (\ref{197}) and found also an analytical expression for the
   coefficient in this equation that reads $-(23/4-4\log 2)\ =
   -1.970$. This result has also been confirmed in recent
   Ref.\cite{sapirstein_03}.

      At this point it is worth returning to the problem of gauge
      invariance.  It is easy to demonstrate by conventional methods
      that the sum of all the diagrams in Fig. \ref{five} is gauge
      invariant. This fact should be compared with the discussion of
      the gauge invariance in Section \ref{equality}, see after
      Eq.(\ref{ddd}).  However, for practical applications one often
      fixes the gauge because this allows one to simplify lengthy
      analytical calculations.  This is the way the calculations were
      performed in
      \cite{k_jpb_02,milstein_sushkov_terekhov_02,sapirstein_03}.
      Since these works employed different gauges, the first one used
      the Feynman gauge while the second and third ones relied on the
      Fried-Yennie gauge (also called the Yennie gauge), their mutual
      agreement provides an additional helpful check of the validity
      of the final result Eq.(\ref{197}).
     
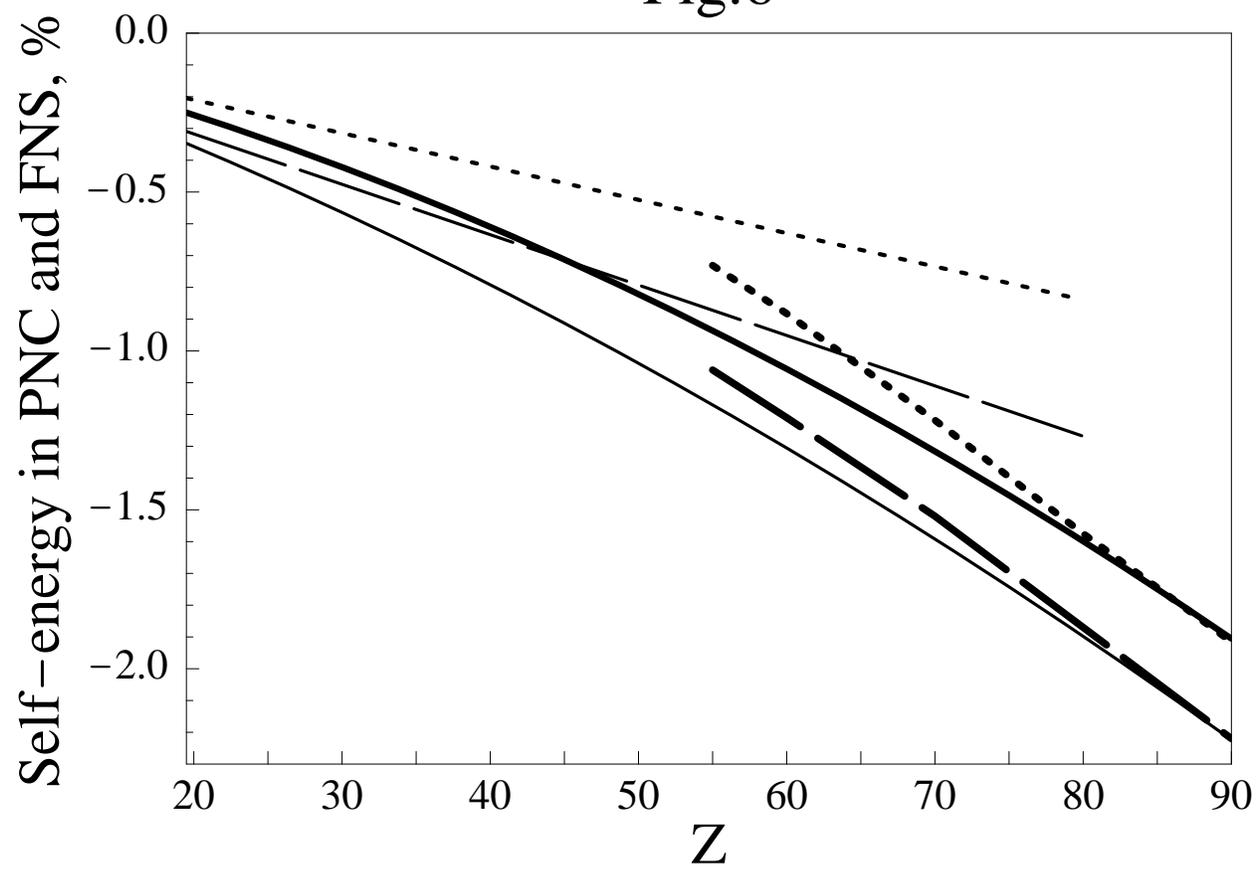
  \begin{figure}[h]
  \caption{\label{six}
    Thick dotted, thin dotted and thick solid lines show e-line
    corrections (self-energy plus vertex) to the PNC amplitude
    predicted by Eq.  (\ref{final}) of \cite{kf_prl_02}, by the linear
    approximation Eq.(\ref{197}) of \cite{k_jpb_02}, and by the
    interpolating Eq.(\ref{interp}) of \cite{k_jpb_02} (all same as in
    Fig.  \ref{seven}).  The thin dashed line, thick dashed line, and
    the dashed-dotted line - the e-line corrections to the FNS energy
    shifts in the linear approximation (\ref{FNS}), from results of
    \cite{blundell_sapirstein_johnson_92}, and the interpolating
    formula (\ref{intFNS}). Note the similarity between the PNC and
    FNS problems.  }\end{figure}

      Eq.(\ref{197}) provides a simple transparent estimate for the
      contribution of the e-line corrections. One point to be noted is
      a strong dependence of the result on the nuclear charge that
      makes the correction large for heavy atoms; remember the
      discussion of this point in Section \ref{intro}.  This behavior
      of the self-energy correction differs qualitatively from what we
      saw for the vacuum polarization in section \ref{vacuum and
        related}, where the higher-orders in the $Z \alpha$-expansion
      were found small. A notable feature in Eq.(\ref{197}) is the
      large coefficient $\sim \! -2.0$ on the right-hand side. {\em
        Naively} one could expect this coefficient to be smaller, of
      the order of $ \sim 1/\pi$. It is interesting that a similar
      ``numerical enhancement'' occurs for the e-line radiative
      correction for the energy shift that is due to the finite
      nuclear size (FNS).  This correction was examined analytically
      in Refs.\cite{pachucki_93,eides_97}, and more recently in Ref.
      \cite{milstein_sushkov_terekhov_02}.  The result can be written
      as

  \begin{eqnarray} \nonumber
  \delta^{ ( \mathrm {e-line},1 )}_{\mathrm {FNS}\,2 } & = &
  -\left( \frac{23}{4}-4\ln 2\right) Z \alpha^2  
\\ \label{FNS}
& = &- 2.978 \,Z\alpha^2 ~,
     \end{eqnarray}
     where the analytical expression was derived in
     \cite{milstein_sushkov_terekhov_02}.  We see that indeed, the
     e-line corrections to the FNS energy shift are governed by the
     large coefficient $\sim \!-3.0$ in (\ref{FNS}), similar to
     (\ref{197}) for the PNC amplitude. Fig. \ref{six} examines this
     similarity in more detail.  It shows data available for relative
     e-line corrections for the two problems mentioned above, namely
     for the PNC amplitude and FNS energy shifts. The linear in $Z$
     approximations (\ref{197}),(\ref{FNS}) (that are valid for
     sufficiently small values of $Z$) are compared in this figure
     with results that follow from numerical calculations of
     Ref.\cite{cheng_93}.  We observe a very close quantitative
     similarity between corrections to PNC and FNS.  In both cases the
     linear approximations (\ref{197}) and (\ref{FNS}) predict large
     negative corrections, which qualitatively agrees with results
     based on numerical calculations for heavy atoms.  The numerical
     validity of the linear approximations seems to be limited by the
     region below $Z = 55$, for higher $Z$ they underestimate the
     effect.  The Cs atom lies on the border, where results of the
     small-$Z$ and large-$Z$ approaches agree reasonably well.

     Numerical data used in Eq.(\ref{ddd}) incorporates an error that
     increases for smaller values of $Z$, see Tables III and IV of
     \cite{cheng_93}.  In order to reduce the impact of this error we can
     combine together Eqs.(\ref{ddd}),(\ref{197}).  One can
     approximate all nonlinear terms omitted in (\ref{197}) by the
     simplest quadratic function and choose the corresponding coefficient
     to reproduce the results based on numerical data for very large
     $Z$, $Z \sim 90$, where the numerical errors are small.  This
     approach gives the following interpolating formula \cite{k_jpb_02} 
     for the corrections to the PNC amplitude

  \begin{equation}\label{interp} \delta^{( \mathrm { e-line,\,int})
  }_{\mathrm {PNC}} = - 1.97 \, Z\alpha^2\, (\, 1+1.55 \,Z \alpha
  \,)~.  
  \end{equation} 
  Fig. \ref{six} shows that the data available for large-Z and small-Z
  regions is very smooth. Therefore {\em any} reasonable interpolation
  would produce a pattern close to Eq.(\ref{interp}).  We can be
  certain therefore that (\ref{interp}) gives reliable numerical data.
  For the $^{133}$Cs atom Eq.(\ref{interp}) predicts $-0.9(1) \% $;
  compare this result with $-0.6 \%$ of (\ref{197}) and $ -0.73(20) \% $
  that follows from Eq.(\ref{ddd}) \footnote{ Ref.  \cite{kf_prl_02},
    where this result was first derived, estimated the uncertainty by
    adopting the error $0.2 \% $ of Ref. \cite{kf_prl_02}. A more
    accurate analysis demonstrates that this estimate is too
    conservative; a more realistic one is $ 0.1 \% $ taken above.}.
  For the Tl atom Eq.(\ref{interp}) predicts $-1.60 \% $ that agrees
  with the prediction of (\ref{ddd}) discussed above (as it should,
  since the data of \cite{cheng_93} incorporates only very small
  errors for heavy atoms).

     It should be noted that Eq.(\ref{ddd}) does not rely on the
     perturbation theory in powers of $Z \alpha$, effectively
     including all nonlinear corrections. This makes it accurate
     even at large values of the nuclear charge. It should therefore
     be considered as a formula that provides a convenient short-cut
     presentation of the nonlinear result.  A similar formula was
     suggested in \cite{k_jpb_02} to reconcile the data for the e-line
     corrections to the energy shifts due to FNS for the $s_{1/2}$
     state.  The linear in $Z$ term of Eq.(\ref{FNS}) and numerical
     data of \cite{cheng_93} available for heavy atoms can be
     interpolated by

     \begin{equation}
       \label{intFNS}
  \delta^{( \mathrm{ e-line,\,int} )}_{\mathrm {FNS} } 
  = - 2.978 \, Z\alpha^2\, 
  (\, 1+0.85 \,Z \alpha  \,)
     \end{equation}
     Fig. \ref{six} shows that this interpolation is reliable.  This
     implies that for relatively light atoms $Z\simeq 55 - 75$
     numerical results for the $s_{1/2}$ energy corrections should be
     amended by slightly larger values.

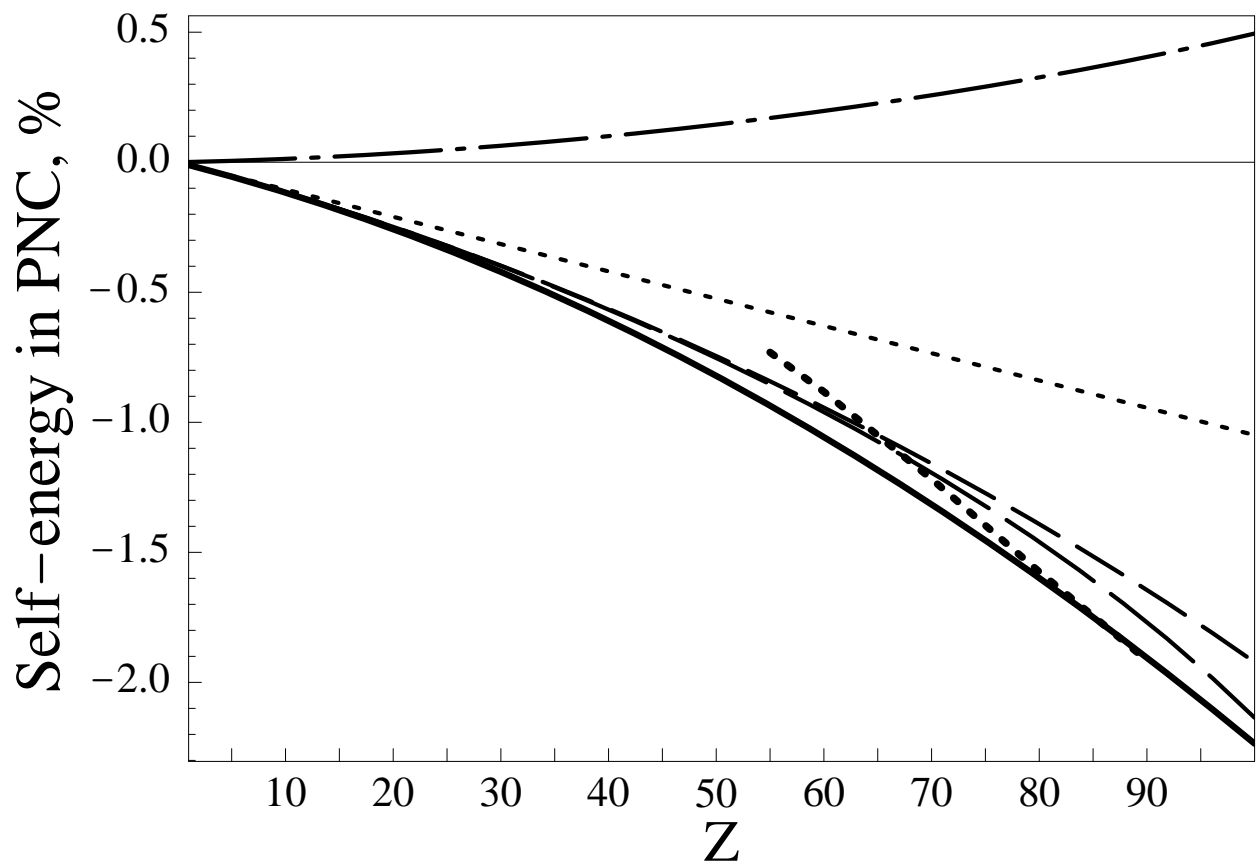
  \begin{figure}[h] \caption{\label{seven} 
      Relative contribution of the e-line corrections (self-energy
      plus vertex) to the PNC amplitude ( \% ) versus the nuclear
      charge.  Thick dotted, thin dotted and thick solid lines show
      predictions of Eq.  (\ref{final}) of \cite{kf_prl_02}, of the
      linear approximation (\ref{197}) of \cite{k_jpb_02}, and the
      interpolating Eq.(\ref{interp}) of \cite{k_jpb_02} respectively,
      as in Fig.  \ref{six}. Dashed-dotted line, thin dashed line and
      thick dashed line - results of Milstein {\em et al} from
      \cite{milstein_sushkov_01,milstein_sushkov_terekhov_02,%
        milstein_sushkov_terekhov_Log_all_orders} respectively. Note
      the convergence of the latest results of the two groups, which
      strongly deviated at the initial stage.  }\end{figure}

     An alternative method to include the nonlinear corrections was
     developed in
     Refs.\cite{milstein_sushkov_terekhov_02,%
       milstein_sushkov_terekhov_Log_all_orders} that relies on direct
     analytical calculations. This approach faces a difficulty because
     even in the next-to-leading order $\sim Z^2\alpha^3$ the
     calculations become complicated.  Ref.\cite{milstein_sushkov_01}
     showed, however, that in this order there appears a large
     logarithmic factor that makes the correction proportional to $\sim
     Z^2\alpha^3 \ln 1/mr_\mathrm{N}$.  Calculations with the
     logarithmic accuracy, when some constant is neglected, are much
     more feasible. The elegant result for the thus calculated
     second-order correction (index 2 in the superscript) was derived
     in \cite{milstein_sushkov_terekhov_02}

   \begin{equation} \label{mstlog}
     \delta_{\mathrm{PNC},sp}^\mathrm{e-line,2} = -\frac{Z^2
     \alpha^3}{\pi} \left( \frac{15}{4}-\frac{\pi^2}{6} \right) \, \ln
     \frac{b}{mr_\mathrm{N} }~, \end{equation} where $b = \exp
     [1/(2\gamma) - C-5/6]$.  The final result of
     \cite{milstein_sushkov_terekhov_02} is given by the sum of the
     linear term Eq.(\ref{197}) and the second order correction
     Eq.(\ref{mstlog}). It is compared in Fig.  \ref{seven} with
     predictions of Eqs. (\ref{ddd}) and (\ref{interp}).  For the sake
     of completeness the figure also shows the previous results of
     Ref.  \cite{milstein_sushkov_01}.  For all values of the nuclear
     charges the results of \cite{milstein_sushkov_terekhov_02} are
     close to our results derived from Eqs.(\ref{ddd}) and
     (\ref{interp}), as was emphasized in \cite{kf_comm_02}. Recently
     Milstein {\em et al}
     \cite{milstein_sushkov_terekhov_Log_all_orders} refined their
     arguments, demonstrating that the logarithmic factor similar to
     the one in Eq.(\ref{mstlog}) exists in all higher order terms of
     the $Z\alpha$-expansion. Ref.
     \cite{milstein_sushkov_terekhov_Log_all_orders} found the
     corresponding contribution of the higher-order terms numerically.
     This latest calculation, also shown in Fig.\ref{seven}, is even
     closer to our results given by Eqs.(\ref{ddd}) and
     (\ref{interp}).  The deviation from (\ref{interp}) is below $ 9
     \% $ for a wide range of atomic charges.  Ref.
     \cite{milstein_sushkov_terekhov_Log_all_orders} claims their
     error to be of the order of $\sim 5 \% $; our estimate for the
     accuracy of Eq.(\ref{ddd}), which limits the error at large $Z$,
     is a {\em few} percent, see Section \ref{equality}.  Within these
     errors the results of the two groups completely agree.  The fact
     that the remaining discrepancy increases for heavy atoms
     indicates probably that it is due to terms of the
     $Z\alpha$-expansion still unaccounted for in
     \cite{milstein_sushkov_terekhov_Log_all_orders}. 
     Overall, Fig.\ref{seven} shows good agreement of the recent results
     obtained by the two groups using different approaches. The sharp
     contradiction that existed during the initial stages of this
     research (compare the dashed-dotted line with the thick-dotted
     line) makes the latest convergence even more satisfying and
     trustworthy .
     
     There is an interesting physical link between the chiral
     invariance identity of Eq.(\ref{ddd}) and the perturbation theory
     approach of Eqs.(\ref{197}),(\ref{mstlog}).  Remember that in
     deriving Eq.(\ref{final}) in the previous section we focused our
     attention at one stage of the consideration at small distances $r
     \le Z\alpha/m$.  The claim was that if this region is proved to
     comply with Eq.(\ref{final}), then the outer region $r \ge
     Z\alpha/m$ would inevitably comply with it as well, see
     discussion after Eq.(\ref{rr}). A similar philosophy lies behind
     the logarithmic approximation in Eq.  (\ref{mstlog}). The
     integration that leads to the logarithmic function in this
     equation is saturated at small distances. The omitted constant is
     related to larger distances, but it is of lesser importance and
     can, in the simplest approximation, be neglected. Thus in both
     approaches the region of small distances proves to be the most
     important.

     This link between the chiral invariance identity and the
     perturbation theory shows that Eq.(\ref{ddd}) should be valid when
     one restricts consideration to the logarithmic approximation.
     This means that the term $\propto Z^2 \alpha^3 \ln mr_\mathrm{N}$
     should exist not only in the PNC problem, see Eq.(\ref{mstlog}),
     but in the FNS problem for the $s_{1/2}$ and $p_{1/2}$ partial
     waves as well.  Eq.(\ref{ddd}) predicts a linear relation between
     the three coefficients of these three terms.
     Ref.\cite{milstein_sushkov_terekhov_02} demonstrated that the three
     coefficients mentioned are all equal, in compliance with
     Eq.(\ref{ddd}). This fact can be considered either as a
     consequence of Eq.(\ref{ddd}), or, alternatively, as its
     independent verification, being fruitful either way.

     The picture of reliability presented above based on the complete
     accord of all available data for the self-energy corrections in
     atoms has recently been put to test {\em again} by Milstein {\em
     et al}.  They have claimed in Ref.
     \cite{milstein_sushkov_terekhov_controversial_Log} that for small
     values of the nuclear charge $Z\simeq 1$ the self-energy
     corrections for the energy shift induced by the FNS in the
     $p_{1/2}$ partial wave are positive and very large, being two
     orders of magnitude greater than anticipated previously for the
     hydrogen atom. For the $p_{3/2}$ state the proposed enhancement
     is even stronger, four orders of magnitude.  This conclusion, if
     correct, could be related to the PNC problem for light elements
     through the equality of \cite{kf_prl_02} mentioned above that
     binds together the radiative corrections for the PNC and FNS
     problems.  It needs, however, to be pointed out that a
     questionable approach is applied in
     \cite{milstein_sushkov_terekhov_controversial_Log}. It uses the
     {\em small}-momenta expansion for the vertex operator. When this
     technique is used for the short-range potential there arises a
     contradiction because the short-range potential needs {\em large}
     momenta.  Attempts to use a similar line of arguments for
     calculations of radiative corrections have been made previously,
     but it was recognized that they strongly overestimate the effect,
     by orders of magnitude.  A necessity to be very careful, to avoid
     using the small-momenta asymptotic dealing with the short-range
     potentials, was clearly stated long time ago by Lepage {\it et
     al} \cite{lepage_81}.  One is forced to presume therefore that
     conclusions of \cite{milstein_sushkov_terekhov_controversial_Log}
     are not convincing (probably they should be even called
     preliminary, if one uses the latter term as in 
     Ref.\cite{milstein_sushkov_01}). Fortunately, these results do
     not influence either the above mentioned good numerical agreement
     that was achieved for heavy atoms, or the analytical agreement
     between the two groups within the $Z\alpha $-expansion that
     defines the PNC in light elements.

     Discussions in this section are almost entirely devoted to
     analytical methods, the only exception was made for the numerical
     data available for corrections to the energy shifts due to the
     FNS. The main reason for this is that direct numerical
     calculations for the self-energy corrections for the PNC
     amplitude have not yet been reported.  Presumably the
     difficulties in this approach are significant, though not being
     experts in this area we would not speculate on the details. We
     wish to mention, however, that developments in this direction are
     quite desirable due to several reasons.  One of them, purely
     theoretical, has come into existence only recently.  Numerical
     studies can determine independently the accuracy of the chiral
     identity (\ref{ddd}).  Knowing the limitations of this relation in
     the PNC problem, one would be able to estimate the accuracy of
     other relations of this type that can be derived and used in
     similar problems in the future.
     
     Summarizing, we discussed above two different methods for
     calculation of the self-energy corrections, one related to the
     chiral invariance identity (\ref{ddd}), the other one based on
     the $Z\alpha$-expansion. They give close results proving that the
     self-energy corrections to the PNC amplitude are large and
     negative. Most importantly, this fact brings the experimental
     data for the $6s - 7s $ PNC amplitude in $^{133}$Cs into
     agreement with the standard model.

     \section{ comparison with experimental
     data }  

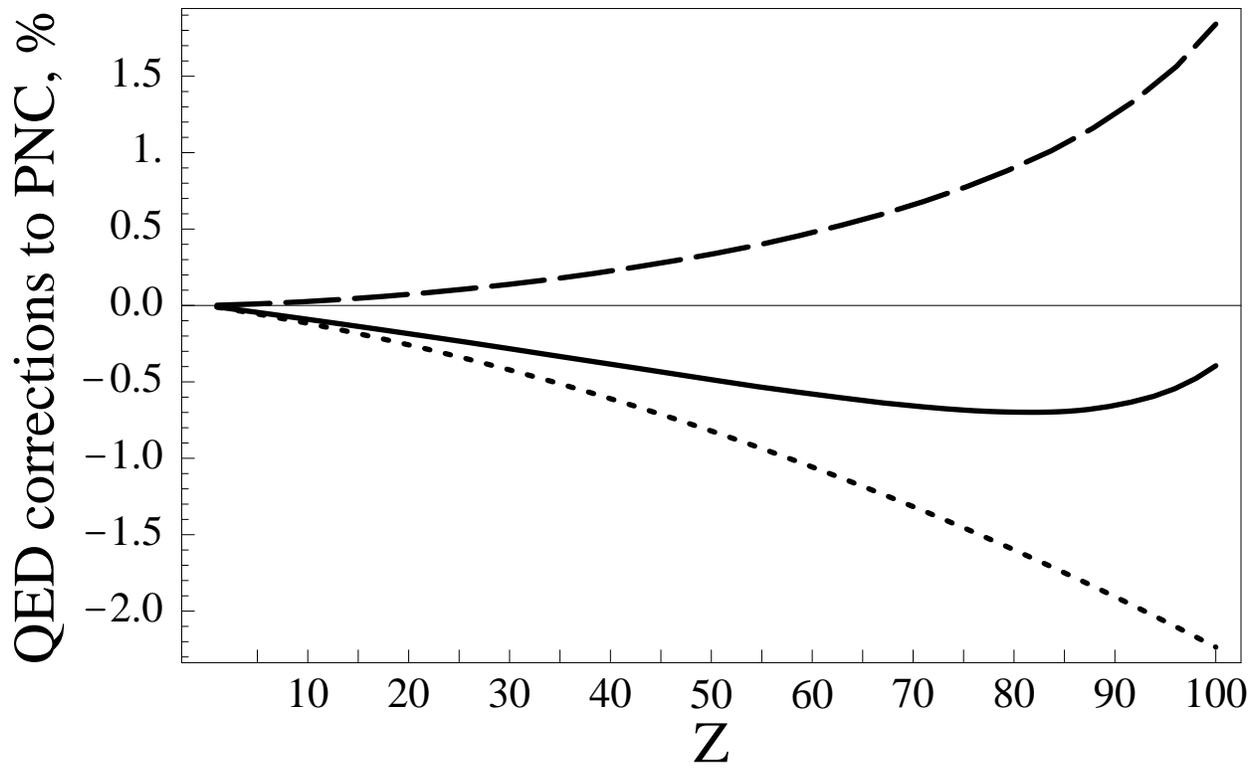
\begin{figure}[h] \caption{\label{eight} 
     QED radiative corrections to the PNC amplitude (\%) versus the
      nuclear charge. The dotted line - vacuum polarization Eq.(\ref{w}),
      thin line - self energy correction Eq.(\ref{interp}), thick line -
      total correction Eq.(\ref{totalQED}).  
}\end{figure}

     Fig. \ref{eight} summarizes the data for the QED radiative
     corrections presenting the vacuum polarization correction
     $\delta^{( \mathrm { VP }) }_{\mathrm {PNC}}$ calculated with the
     help of Eq.(\ref{w})-(\ref{w2}) and the self-energy (plus vertex)
     radiative correction $ \delta^{( \mathrm { e-line }) }_{\mathrm
     {PNC}}$ described by Eq.(\ref{interp}), as well as the total
     radiative correction \begin{equation} \label{totalQED} \delta^
     {( \mathrm { tot }) }_{\mathrm {PNC}} = \delta^{( \mathrm { VP })
     }_{\mathrm {PNC}} + \delta^{( \mathrm { e-line }) }_{\mathrm
     {PNC}}.  
   \end{equation} 
   We see that both the vacuum polarization and the self-energy
   corrections rise quickly with the nuclear charge, but their
   opposite signs make the total correction smoother.  We find
   $\delta^{( \mathrm { tot }) }_{\mathrm {PNC}} =-0.54 \% $ for the
   Cs atom and $\delta^{( \mathrm { tot }) }_{\mathrm {PNC}}= -0.70 \%
   $ for Pb,Tl and Bi.

     Let us discuss the implications of these results, first  for the
     $6s-7s$ PNC amplitude in $^{133}$Cs. The standard model value for
     the nuclear weak charge for Cs \cite{hagivara_02} is

  \begin{equation}\label{QW}
  Q_W (^{133}{\mathrm Cs}) = \,-73.09 \,\pm\,(0.03)~.
     \end{equation}
     Ref. \cite{dzuba_02} refined previous calculations of Ref.
     \cite{dzuba_89}, extracting from the experimental PNC amplitude of
     Ref. \cite{wood_97} the weak charge

  \begin{eqnarray}\label{72.45}
&&  Q_W ^{(\mathrm{C+B+N}) }(^{133}{\mathrm Cs}) = 
\\ \nonumber
&&-72.45\pm(0.29)_{\mathrm {expt} }\pm(0.36)_{\mathrm   {theor}  },
     \end{eqnarray}
     with the theoretical error $0.5\%$. This value includes the
     correlation and the Breit corrections, as well as the neutron
     skin corrections (superscript C+B+N), but does not take into
     account the radiative corrections \footnote{In order to compare
       Eq.(\ref{72.45}) with results of Ref.\cite{dzuba_02} one needs
       to extract from -72.45 the vacuum polarization correction $0.4
       \% $. This results in $Q_W ( ^{133}{\mathrm Cs} ) = -72.16$ in
       agreement with Eq.(43) of \cite{dzuba_02}.}.  The neutron
     nuclear radius is slightly larger than its proton radius. This
     provides a possibility for the electron to interact with neutrons
     in the region outside of the proton core, where the wave function
     is slightly smaller than inside the nucleus. The corresponding
     reduction of the PNC amplitude is called the neutron skin
     correction in PNC.  According to estimates of Derevianko
     \cite{derevianko_02} the neutron skin correction is
     approximately $-0.2 \% $ for Cs. The calculations of
     Ref.\cite{dzuba_02} took this effect into account.
     
     Eq.(\ref{72.45}) is consistent with $Q_W(^{133}{\mathrm Cs}) =
     \,-72.21 \,\pm\,(0.28)_{\mathrm {expt}}\,\pm\, (0.34)_{\mathrm
       {theor}}$ that was adopted in \cite{johnson_01} by taking the
     average of the results of
     Refs.\cite{dzuba_89,blundell_90,kozlov_01}, and accepting the
     theoretical error $0.4\%$ proposed in \cite{bennett_wieman_99}.
     This value for the weak charge includes the vacuum polarization
     correction $0.4 \%$. Extracting it and taking into account $-0.2
     \% $ for the nuclear skin correction one obtains $-72.35$ that is
     close to Eq.(\ref{72.45}).
     
     The radiative corrections derived from results presented in Fig.
     \ref{eight} are $-0.54\pm(0.10) \%$; the error reflects the
     uncertainty of the self-energy radiative correction.
     Eq.(\ref{72.45}) combined with the QED radiative corrections gives

  \begin{eqnarray}\label{72.84}  
&&  Q_W^{\mathrm {Total}}(^{133}{\mathrm Cs}) =  
\\
\nonumber
&&  \,-72.84 \pm(0.29)_{\mathrm {expt}}\pm(0.36)_{\mathrm {theor}}~.
     \end{eqnarray}
     where Total = Correlations + Breit corrections + Neutron skin
     correction + QED radiative corrections.  The agreement with the
     standard model given in Eq.(\ref{QW}) is good.  It is so good, in
     fact, that we hasten to remind the reader a historical aspect of
     the problem.  The theory has come up with Eq.(\ref{72.84}) after
     a turbulent period of research during which the experimental data
     were widely anticipated to be in contradiction with the standard
     model.  This makes the found ``unexpected''agreement more
     objective.
     
     Let us consider now the case of the thallium atom that was
     studied experimentally in
     \cite{berkeley_79,berkeley_81,berkeley_85,oxford_91_Tl,oxford_95,
     seattle_95}, the calculations were performed in
     \cite{Tl_87,kozlov_pra_01}, that give close results for the
     many-electron correlation. The corrections to the result of
     \cite{Tl_87} should include contributions of the Breit
     interaction -0.98\% \cite{0.98dzuba}.  As was mentioned, the QED
     radiative corrections for Tl give $-0.70 \% $. If we adopt
     the neutron skin correction $-0.2 \% $, then the total
     theoretical result reads

  \begin{equation}
    \label{qtl}
\!\! R= \mathrm{Im }( E_{\mathrm{PNC}}/M1 )= 
  -15.64 (-Q_W/N) \! \cdot \! 10 ^{-8}.
  \end{equation}
  Comparing with the measured value of Ref.\cite{seattle_93} that
  predicts $R=-14.68(17)\cdot 10^{-8}$, we obtain the weak charge for
  $^{205}$Tl

   \begin{equation}
     \label{qwtk}
  Q_W(^{205}\mathrm{Tl})=-116.4 \pm
   (1.3)_\mathrm{expt} \pm (3.4)_\mathrm{theor}
   \end{equation}
   in good agreement with the standard model prediction
\cite{hagivara_02}

   \begin{equation} \label{Tl_sm} Q_W(^{205}\mathrm{Tl})= -116.7(1) ~.
     \end{equation}

  \section{conclusion}
     \label{conclusion}
     
     The QED corrections for parity nonconservation have recently
     emerged as an important ingredient in the analyses of modern
     experimental data.  The improvement in the experimental accuracy,
     particularly for the $6s-7s$ PNC amplitude in $^{133}$Cs, and the
     progress of atomic structure calculations revealed a possible gap
     between the experiment data and atomic structure calculations on
     one side and the standard model on the other.
     
     The QED corrections, which embrace the Breit corrections and the
     radiative corrections, reconcile the experimental data with the
     standard model. The area has evolved and developed very rapidly.
     Over the past couple of years, even over a year, the situation
     has changed dramatically.  A short time ago one could be
     seriously contemplating the possibility to modify the standard
     model to accommodate a possible deviation of atomic experimental
     data with the standard model. Presently it has become clear that
     modifications of the model are to be postponed, one simply needs
     to calculate everything accurately.  {\it Everything} here is
     essential. Different parts of the QED corrections are all very
     large and have different signs. One must be certain that all
     parts are included and accounted for properly. That is why we
     spent some time and space above presenting and comparing all
     available results for the radiative corrections. The conclusion
     is that {\em all} data for the radiative corrections fit together
     well. For each possible correction several different methods of
     calculation were employed by different groups with the same final
     conclusions.  The self-energy corrections, that proved to be by
     far the most difficult ones, are known presently with an
     uncertainty of better than 10 \%, which is sufficient for the
     present day experimental accuracy. One can contemplate
     a significant reduction of this error, if required.
     
     The calculations of the radiative corrections for PNC were based
     on vast expertise accumulated in other related
     research areas that include the Lamb shift, the hyperfine
     interaction, the nuclear finite-size correction. In turn, some
     methods of calculations that have recently evolved in relation to
     PNC may find applications in other areas. One of them is the chiral
     invariance identity Eq.(\ref{ddd}), that was originally applied
     to express radiative corrections to PNC in terms of the
     corrections to energy shifts due to the finite nuclear size.
     Similar identities can be derived for other related problems.
     One can also mention a simple convenient expression that gives the
     variation of the electron wave function at the atomic nucleus in
     terms of the two first moments of the perturbative potential
     Eq.(\ref{final}).
     
     In conclusion, large QED radiative corrections to the parity
     nonconservation amplitude in heavy atoms reconcile the
     accurate atomic experimental data on parity nonconservation for
     the $6s-7s$ PNC amplitude in $^{133}$Cs of Wood {\em et al}
     \cite{wood_97} with the standard model.

     
     Discussions with D.Budker, J.S.M.Ginges, W.Greiner, G.F.Gribakin,
     M.G.Kozlov, V.M.Shabaev, and A.V.Solov'yov are greatly
     appreciated.  This work was supported by the Australian Research
     Council.




\clearpage
\pagebreak

\nopagebreak[4]

\nopagebreak[4]

  \vspace{1cm}

\pagebreak

\nopagebreak[4]

\begin{thebibliography}{99}



  \bibitem{bouchiat_74}
    
    M.A.Bouchiat and C.Bouchiat J.Phys.  (Paris) {\bf 35}, 899 (1974);
    {\bf 36}, 493 (1974).


  \bibitem{bouchiat_82}
    
    M.A. Bouchiat, J. Gu{\'e}na, L. Hunter, and L. Pottier, Phys.
    Lett. B {\bf 117}, 358 (1982); errata {\bf 121}, 456 (1983).

    
  \bibitem{bouchiat_84}
  
    M.A. Bouchiat, J. Gu{\'e}na, L. Pottier, and L. Hunter, Phys.
    Lett.  B {\bf 134}, 463 (1984).
  
  \bibitem{bouchiat_85}
  
    M.A. Bouchiat, J. Gu{\'e}na, and L. Pottier, J. Phys. (Paris) {\bf
      46}, 1897 (1985).

  \bibitem{bouchiat_86}
    
    M.A. Bouchiat, J. Gu{\'e}na, and L. Pottier, J. Phys. (Paris)
    {\bf 47}, 1175 (1986).


  \bibitem{bouchiat_86_1}
      
      M.A. Bouchiat, J.  Gu{\'e}na, L. Pottier, and L. Hunter, J.
      Phys. (Paris) {\bf 47}, 1709 (1986).
  

  \bibitem{bouchiat_02}
  
    J. Gu{\'e}na, D. Chauvat, Ph. Jacquier, E. Jahier, M. Lintz, A.V.
    Papoyan, S. Sanguinetti, D. Sarkisyan, A. Wasan, and M.A.
    Bouchiat, physics/0210069.


    
  \bibitem{boulder_85}
  
    S.L. Gilbert, M.C. Noecker, R.N. Watts, and C.E. Wieman, Phys.
    Rev.  Lett. {\bf 55}, 2680 (1985).


  \bibitem{boulder_86}
    
    S.L.  Gilbert and C.E. Wieman, Phys.  Rev. A {\bf 34}, 792 (1986).

    
  \bibitem{boulder_88}
  
    M.C. Noecker, B.P. Masterson, and C.E. Wieman, Phys. Rev. Lett.
    {\bf 61}, 310 (1988).
  
  \bibitem{wood_97}
  
    C.S. Wood, S.C. Bennett, D. Cho, B.P. Masterson, J.L. Roberts,
    C.E.  Tanner, and C.E. Wieman, Science {\bf 275}, 1759 (1997).


    
  \bibitem{berkeley_79}
  
    R. Conti, P. Bucksbaum, S. Chu, E. Commins, and L. Hunter, Phys.
    Rev.  Lett. {\bf 42}, 343 (1979).
  
  \bibitem{berkeley_81}
  
    P.K. Bucksbaum, E.D. Commins, and L.R. Hunter, Phys.  Rev. Lett.
    {\bf 46}, 640 (1981);
  
    P.H. Bucksbaum, E.D. Commins, and L.R. Hunter, Phys. Rev.  D {\bf
      24}, 1134 (1981).

  
  \bibitem{berkeley_85}
  
    P.S. Drell and E.D. Commins,  Phys. Rev. Lett. {\bf 53}, 968
    (1984);
    
    P.S. Drell and E.D. Commins, Phys. Rev. A {\bf 32}, 2196 (1985).




    
  \bibitem{oxford_91_Tl}
    
    T.D. Wolfenden, P.E.G. Baird, and P.G.H. Sandars, Europhys. Lett.
    {\bf 15}, 731 (1991).

  \bibitem{oxford_95}
    
    N.H. Edwards, S.J. Phipp, P.E.G. Baird, and S. Nakayama, Phys.
    Rev. Lett. {\bf 74}, 2654 (1995).


  \bibitem{seattle_95}    
    
    P.A. Vetter, D.M. Meekhof, P.K. Majumder, S.K. Lamoreaux, and E.N.
    Fortson, Phys. Rev. Lett. {\bf 74}, 2658 (1995).
    
    
    

    
  \bibitem{seattle_83}
    
    T.P. Emmons, J.M. Reeves, and E.N. Fortson, Phys. Rev. Lett. {\bf
      51}, 2089 (1983); errata {\bf 52}, 86 (1984).
    
  \bibitem{seattle_93}
    D.M. Meekhof, P. Vetter, P.K. Majumder, S.K. Lamoreaux, and E.N.
    Fortson, Phys. Rev. Lett. {\bf 71}, 3442 (1993).


  \bibitem{oxford_96}
    
    S.J. Phipp, N.H. Edwards, P.E.G. Baird, and S. Nakayama, J. Phys.
    B {\bf 29}, 1861 (1996).



    
  \bibitem{novosibirsk_78} 
    
    L.M. Barkov and M.S. Zolotorev, Pis'ma Zh.  Eksp. Teor. Fiz. {\bf
      27}, 379 (1978) [JETP Lett. {\bf 27}, 357 (1978)].

  \bibitem{novosibirsk_78_1} 
    
    L.M. Barkov and M.S. Zolotorev, Pis'ma Zh. Eksp. Teor.  Fiz. {\bf
      28}, 544 (1978) [JETP Lett. {\bf 28}, 503 (1978)].
    
  \bibitem{novosibirsk_79} 
    
    L.M. Barkov and M.S. Zolotorev, Phys.  Lett. B {\bf 85}, 308
    (1979).

  \bibitem{novosibirsk_80} 
    
    L.M. Barkov and M.S. Zolotorev, Zh. Eksp. Teor. Fiz. {\bf 79}, 713
    (1980) [Sov. Phys. JETP {\bf 52}, 360 (1980)].

 
  \bibitem{moscow_84}
    
    G.N. Birich, Yu.V. Bogdanov, S.I. Kanorskii, I.I. Sobel'man, V.N.
    Sorokin, I.I. Struk, and E.A. Yukov, Zh. Eksp. Teor. Fiz. {\bf
      87}, 776 (1984) [Sov. Phys. JETP {\bf 60}, 442 (1984)].


  \bibitem{oxford_87_5/2}
    
    J.D. Taylor, P.E.G. Baird, R.G. Hunt, M.J.D. Macpherson, G.
    Nowicki, P.G.H. Sandars, and D.N. Stacey, J. Phys. B {\bf 20},
    5423 (1987).


  \bibitem{oxford_93}
    
    R.B. Warrington, C.D. Thompson, and D.N. Stacey, Europhys. Lett.
    {\bf 24}, 641 (1993).



    
  \bibitem{seattle_81}
    
    J.H. Hollister, G.R. Apperson, L.L. Lewis, T.P. Emmons, T.G. Vold,
    and E.N. Fortson,
    Phys. Rev. Lett. {\bf 46}, 643 (1981).


    
  \bibitem{oxford_87_3/2}
    
    M.J.D. Macpherson, D.N. Stacey, P.E.G. Baird, J.P. Hoare, P.G.H.
    Sandars, K.M.J. Tregidgo, and Wang Guowen, Europhys. Lett. {\bf
      4}, 811 (1987).
    
  \bibitem{oxford_91_Bi}
  
    M.J.D. Macpherson, K.P. Zetie, R.B. Warrington, D.N. Stacey, and
    J.P. Hoare, Phys. Rev. Lett. {\bf 67}, 2784 (1991).


    
  \bibitem{khriplovich_91} 
    
    I.B. Khriplovich. {\it Parity Nonconservation in Atomic
      pnenomena.}  (Gordon and Breach, Philadelphia, 1991).


  \bibitem{dzuba_89}
    
    V.A.Dzuba, V.V.Flambaum, and O.P.Sushkov, Phys.  Lett A {\bf 141},
    147 (1989).


  \bibitem{blundell_90}

    S.A.Blundell, W.R.Johnson, and J.Sapirstein, Phys.  Rev. Lett. {\bf
      65}, 1411  (1990);

    S.A.Blundell, J.Sapirstein, and W.R.Johnson, Phys.  Rev. D {\bf 45},
    1602 (1992).


  \bibitem{bennett_wieman_99} 

    S.C.Bennett and C.E.Wieman, Phys. Rev.  Lett.  {\bf 82}, 2484
    (1999); {\bf 82}, 4153 (1999); {\bf 83}, 889 (1999).


  \bibitem{kozlov_01}

    M.G.Kozlov, S.G.Porsev, and I.I.Tupitsyn, Phys. Rev.  Lett. {\bf
      86}, 3260 (2001).

 
  \bibitem{dzuba_01}

    V.A.Dzuba, V.V.Flambaum, and J.S.M. Ginges, hep-ph/0111019.


  \bibitem{dzuba_02}

V.A. Dzuba, V.V. Flambaum, J.S.M. Ginges.  Phys. Rev. D66, 076013
(2002); hep-ph/0204134.

  \bibitem{derevianko_00}

    A.Derevianko, Phys. Rev. Lett. {\bf 85}, 1618 (2000).


  \bibitem{dzuba_harabati_01}

    V.A.Dzuba, C.Harabati, W.R.Johnson, and M.S.Safronova, Phys. Rev A
    {\bf 63}, 044103 (2001).


  \bibitem{sushkov_01}

    O.P.Sushkov, Phys. Rev. A, {\bf 63}, 042504 (2001).


  \bibitem{johnson_01}

    W.R.Johnson, I.Bednyakov, and G.Soff, Phys.  Rev.  Lett. {\bf 87},
    233001-1 (2001).


  \bibitem{milstein_sushkov_01}

    A.I.Milstein and O.P.Sushkov, Phys.Rev. A {\bf 66}, 022108/1-4
    (2002); hep-ph/0109257.


  \bibitem{kf_jpb_02}

    M.Yu.Kuchiev and V.V.Flambaum, J.Phys.B:At.Mol.Opt.Phys. {\bf 35},
    4101 (2002); hep-ph/0205012.


  \bibitem{marciano_sirlin_83}

    W.J.Marciano and A.Sirlin, Phys. Rev. D {\bf 27}, 552 (1983).


  \bibitem{lynn_sandars_94}

    B.W.Lynn and P.G.H.Sandars, J. Phys. B {\bf 27}, 1469 (1994).


  \bibitem{blundell_97}
  S.A.Blundell, K.T.Cheng, and J.Sapirstein, Phys.  Rev. A {\bf 55}, 1857
  (1997).



  \bibitem{sunnergren_98}

    P.Sunnergren, H.Persson, S.Salomonson, S.M.Sneider, I.Lindgren, and
    G.Soff, Phys. Rev.  A {\bf 58}, 1055 (1998).


  \bibitem{kf_prl_02}

    M.Yu.Kuchiev and V.V.Flambaum, Phys.Rev.Lett. {\bf 89} 283002
    (2002); hep-ph/0206124.


  \bibitem{johnson_soff_85}

    W.R.Johnson and G.Soff, At. Data Nuc. Data Tables {\bf 33}, 405
    (1985).


  \bibitem{blundell_92} 

    S.A.Blundell, Phys. Rev. A {\bf 46}, 3762 (1992).


  \bibitem{blundell_sapirstein_johnson_92} 

    S.A.Blundell, J.Sapirstein, and W.R.Johnson, Phys.  Rev. D {\bf 45},
    1602 (1992).


  \bibitem{cheng_93}

    K.T.Cheng, W.R.Johnson and J.Sapirstein, Phys. Rev A {\bf 47}, 1817
    (1993).


  \bibitem{lindgren_93}

    I.Lindgren, H.Persson, S.Salomonson, and A.Ynnerman, Phys. Rev. A
    {\bf 47}, 4555 (1993).


  \bibitem{cheng_02}

    K.T.Cheng. Private communication (2002).


  \bibitem{k_jpb_02}

    M.Yu.Kuchiev, J.Phys.B:At.Mol.Opt.Phys. {\bf 35} 4101 (2002);
    hep-ph/0208196.



  \bibitem{milstein_sushkov_terekhov_02}

    A.I.Milstein, O.P.Sushkov, and I.S.Terekhov, Phys.Rev.Lett.  {\bf
      89} 28003 (2002); hep-ph/0208227.



  \bibitem{kf_comm_02}

    M.Yu.Kuchiev and V.V.Flambaum, hep-ph/0209052.


  \bibitem{milstein_sushkov_terekhov_Log_all_orders}

    A.I.Milstein, O.P.Sushkov, and I.S.Terekhov, hep-ph/0212072.


  \bibitem{milstein_sushkov_terekhov_controversial_Log}

    A.I.Milstein, O.P.Sushkov, and I.S.Terekhov, hep-ph/0212018


  \bibitem{lepage_81}

    G.P.Lepage, D.R.Yenni, and W.Erickson, Phys. Rev. Lett. {\bf 47},
    1640 (1981).


    


   \bibitem{LLIII}
   L.D. Landau  and E.M. Lifshitz   1977
   {\it Quantum mechanics : non-relativistic theory}
   (Oxford, New York, Pergamon Press)



   \bibitem{uehling}

    E.A.Uehling,  Phys.Rev. {\bf 48}, 55 (1935).


 
  \bibitem{fullerton_rinker}

  L.W.Fullerton and J.A.Rinker, Jr., Phys.Rev.A {\bf 13}, 1283 (1976).

 
  \bibitem{wichmann-kroll}

  E.H.Wichmann and N.M.Kroll  {\it Phys.Rev.} {\bf 101}, 843 (1956)


  \bibitem{milstein_strakhovenko_83}


  A.I.Milstein  and V.M.Strakhovenko  
   {\it Sov.ZhETF} {\bf 84}, 1247 (1983).


 \bibitem{df_03}
    V.A.Dzuba and V.V.Flambaum (to be published).




  \bibitem{pachucki_93}

    K.Pachucki, Phys.  Rev. A {\bf 48}, 120 (1993).


  \bibitem{eides_97}

   M.I.Eides and H.Grotch, Phys.  Rev. A {\bf 56}, R2507 (1997);

   M.I.Eides, H.Grotch, and V.A.Shelyuto, Phys. Rep.  {\bf 342}, 63
   (2001).


  \bibitem{history} 
    
    The work \cite{k_jpb_02} was inspired, not to say imposed on us,
    by a sceptical referee of \cite{kf_prl_02} who challenged us to
    support our results using perturbation theory.
    Ref.\cite{k_jpb_02} obliged dutifully, being written as a reply to
    the referee comment. The agreement between
    Refs.\cite{kf_prl_02,k_jpb_02} made our claim stronger, prompting
    us to acknowledge a fruitful, even if unnintentional, impact of our
    reluctant referee.


    \bibitem{sapirstein_03}

J.Sapirstein, K.Pachucki, A.Vietia, and K.T.Cheng, hep-ph/0302202.

  \bibitem{hagivara_02}
K.Hagivara {\it et al}, Phys.Rev. D {\bf 64}, 010001 (2002).


   \bibitem{0.98dzuba}
     
     V.A.Dzuba, privet communication.


  \bibitem{derevianko_02}
  A.Derevianko,  Phys.Rev. A {\bf 65}, 012106 (2002).


  \bibitem{Tl_87} 
    
    V.A. Dzuba, V.V. Flambaum, P.G. Silvestrov, and O.P.  Sushkov.
    J.Phys.  B {\bf 20}, 3297 (1987).


  \bibitem{kozlov_pra_01} 

    M.G. Kozlov, S.G. Porsev, and W.R. Johnson.  Phys. Rev. A {\bf 64},
    052107 (2001).


\end{thebibliography}
  
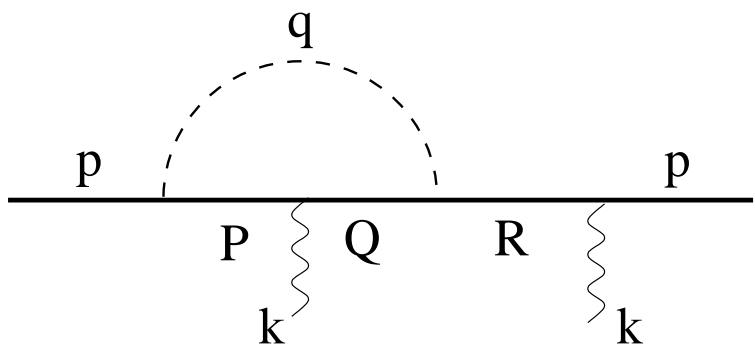
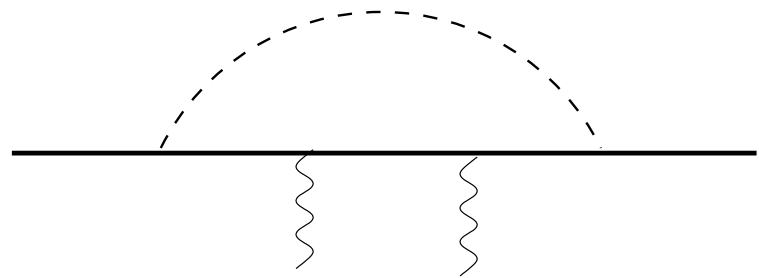
\end{document}